\documentclass[twocolumn,tighten,times]{aastex63}

\usepackage{amsmath}
\usepackage{amssymb}
\usepackage{float}
\usepackage{lipsum}

\usepackage{graphicx}
\usepackage{graphics}
\usepackage{hyperref}

\graphicspath{{./}{figures/}}
\received{November 30, 2020}
\revised{February 26, 2021}
\accepted{March 11, 2021}
\submitjournal{ApJ}

\shorttitle{Variable White Dwarfs from {\em Gaia} and ZTF}
\shortauthors{Guidry et al.}

\begin{document}
\title{I Spy Transits and Pulsations: Empirical Variability in White Dwarfs Using {\em Gaia} and the Zwicky Transient Facility}

\correspondingauthor{Joseph A. Guidry}
\email{josephguidry@utexas.edu}

\author[0000-0001-9632-7347]{Joseph A. Guidry}
\affiliation{Department of Astronomy, The University of Texas at Austin, Austin, TX 78712, USA}

\author[0000-0002-0853-3464]{Zachary P. Vanderbosch}
\affiliation{Department of Astronomy, The University of Texas at Austin, Austin, TX 78712, USA}
\affiliation{McDonald Observatory, Fort Davis, TX 79734, USA}

\author[0000-0001-5941-2286]{J.~J. Hermes}
\affiliation{Department of Astronomy, Boston University, 725 Commonwealth Ave., Boston, MA 02215, USA}

\author[0000-0002-8558-4353]{Brad N. Barlow}
\affiliation{Department of Physics, High Point University, High Point, NC 27268, USA}

\author[0000-0002-0009-409X]{Isaac D. Lopez}
\affiliation{Wentworth Institute of Technology, 550 W Huntington Ave., Boston, MA 02215, USA}
\affiliation{Department of Astronomy, Boston University, 725 Commonwealth Ave., Boston, MA 02215, USA}

\author[0000-0002-2600-7513]{Thomas M. Boudreaux}
\affiliation{Department of Physics and Astronomy, Dartmouth College, Hanover, NH 03755, US}

\author[0000-0002-2764-7248]{Kyle A. Corcoran}
\affiliation{Department of Astronomy, University of Virginia, Charlottesville, VA 22904, USA}

\author[0000-0002-0656-032X]{Keaton J. Bell}
\affiliation{DIRAC Institute, Department of Astronomy, University of Washington, Seattle, WA 98195, USA}
\affiliation{NSF Astronomy and Astrophysics Postdoctoral Fellow}

\author[0000-0002-6748-1748]{M. H. Montgomery}
\affiliation{Department of Astronomy, The University of Texas at Austin, Austin, TX 78712, USA}
\affiliation{McDonald Observatory, Fort Davis, TX 79734, USA}

\author[0000-0003-3868-1123]{Tyler M. Heintz}
\affiliation{Department of Astronomy, Boston University, 725 Commonwealth Ave., Boston, MA 02215, USA}

\author{Barbara G. Castanheira}
\affiliation{Department of Physics, Baylor University, Waco, TX 76798, USA}

\author[0000-0003-1862-2951]{Joshua S. Reding}
\affiliation{Department of Physics and Astronomy, University of North Carolina at Chapel Hill, Chapel Hill, NC 27599, USA}

\author[0000-0002-1086-8685]{Bart H. Dunlap}
\affiliation{Department of Astronomy, The University of Texas at Austin, Austin, TX 78712, USA}

\author[0000-0003-0181-2521]{D. E. Winget}
\affiliation{Department of Astronomy, The University of Texas at Austin, Austin, TX 78712, USA}
\affiliation{McDonald Observatory, Fort Davis, TX 79734, USA}

\author{Karen I. Winget}
\affiliation{Department of Astronomy, The University of Texas at Austin, Austin, TX 78712, USA}
\affiliation{McDonald Observatory, Fort Davis, TX 79734, USA}

\author{J. W. Kuehne}
\affiliation{McDonald Observatory, Fort Davis, TX 79734, USA}

%%%%%%%%%%%%%%%%%%%%%%%%%%%%%%%%%%%%%%%%%%%%%%%%%%%%%%%%%%%%%%%%%%%%%%%

\begin{abstract}

We present a novel method to detect variable astrophysical objects and transient phenomena using anomalous excess scatter in repeated measurements from public catalogs of {\em Gaia} DR2 and Zwicky Transient Facility (ZTF) DR3 photometry. We first provide a generalized, all-sky proxy for variability using only {\em Gaia} DR2 photometry, calibrated to white dwarf stars. To ensure more robust candidate detection, we further employ a method combining {\em Gaia} with ZTF photometry and alerts. To demonstrate the efficacy, we apply this latter technique to a sample of roughly 12${,}$100 white dwarfs within 200\,pc centered on the ZZ Ceti instability strip, where hydrogen-atmosphere white dwarfs are known to pulsate. Through inspecting the top 1\% samples ranked by these methods, we demonstrate that both the {\em Gaia}-only and ZTF-informed techniques are highly effective at identifying known and new variable white dwarfs, which we verify using follow-up, high-speed photometry. We confirm variability in all 33 out of 33 (100\%) observed white dwarfs within our top 1\% highest-ranked candidates, both inside and outside the ZZ Ceti instability strip. In addition to dozens of new pulsating white dwarfs, we also identify five white dwarfs highly likely to show transiting planetary debris; if confirmed, these systems would more than triple the number of white dwarfs known to host transiting debris.

\end{abstract}


%%%%%%%%%%%%%%%%%%%%%%%%%%%%%%%%%%%%%%%%%%%%%%%%%%%%%%%%%%%%%%%%%%%%%%%%%%%%%%%%%%%

                                %INTRODUCTION                
                
%%%%%%%%%%%%%%%%%%%%%%%%%%%%%%%%%%%%%%%%%%%%%%%%%%%%%%%%%%%%%%%%%%%%%%%%%%%%%%%%%%%

\section{Introduction} \label{sec:intro}

As astronomers prepare for the Vera C. Rubin Observatory and its Legacy Survey of Space and Time (LSST, \citealt{2019ApJ...873..111I}), we are entering an exciting era of big-data astronomy, with a rapid increase in synoptic photometric surveys that cover large areas of the sky. This shift is enabling astronomers to discover an exponentially increasing number of stars that go bump in the night.

Our understanding of white dwarf stars, which mark the endpoints of low- and intermediate-mass stars and their planetary systems, will benefit significantly from this new era \citep[e.g.,][]{Fantin2020}. The vast majority of white dwarfs are photometrically constant and make for excellent flux standards \citep{2017MNRAS.468.1946H}, but an important subset of these stellar remnants show photometric variability caused by a range of phenomena, including pulsations, binarity, surface inhomogeneities that rotate in and out of view, and more recently transits by both planets and planetary debris \citep{Vanderburg2015,Vanderbosch2020,2020Natur.585..363V}. Studying photometric variations in white dwarfs can therefore provide much insight into the end stages of stars and planets. For example, the detection and characterization of these variations can enable the exploration of white dwarf interiors by performing asteroseismology on the pulsations present \citep[e.g.,][]{1994ApJ...430..839W}, or constrain the grain size and circumstellar dust properties of systems showing transiting debris \citep[e.g.,][]{2017MNRAS.469.3213H,2019AJ....157..255X}.

Searches for new pulsating white dwarfs have traditionally been focused on a narrow range in photometric color or spectroscopically determined effective temperature  \citep[e.g.,][]{2004ApJ...607..982M,Vincent2020}. This is motivated by the fact that, as white dwarf stars cool monotonically throughout their lifetimes, they will eventually arrive at the ZZ Ceti instability strip, where hydrogen-atmosphere (DA) white dwarfs develop a deep enough convection zone to drive observable pulsations \citep{2015ApJ...812..167G}. ZZ Cetis typically exhibit optical variability with peak-to-peak amplitudes of $1-30\%$ and periods ranging from $100-1500$\,s \citep{Mukadam2013,Bognar2020}.

Prior approaches to mining all-sky surveys for variable white dwarfs relied primarily on searches for coherent variability, manifest as significant peaks in the periodograms of these noise-dominated, sporadically sampled observations \citep[e.g.,][]{2019MNRAS.486.4574R,2020ApJS..249...18C,Coughlin2020}. These studies used variability metrics established upon finding significant periodicities in the Lomb-Scargle periodogram \citep[][see also \citealt{VanderPlas2018}]{1976Ap&SS..39..447L, 1982ApJ...263..835S} computed from each object's time-series photometry. Therefore, aperiodic photometric variations have so far mostly been neglected, despite the fact that the first pulsating white dwarf, HL Tau 76, was discovered from anomalous point-to-point scatter when observed for use as a flux standard \citep{1968ApJ...153..151L}. Similarly, {\em Gaia} DR2 catalogs of variables have mostly neglected short-term variable objects such as ZZ Cetis, since established periodic variability was required to flag a source as variable in {\em Gaia} DR2 \citep{Holl2018}.

Searches for other types of variability in white dwarfs have been less focused. Astronomers have unsuccessfully searched for transits from close-in planets or debris for more than a decade using large samples and wide-field surveys (e.g., \citealt{2011MNRAS.410..899F,2014ApJ...796..114F,2018MNRAS.474.4603V,2019MNRAS.486.4574R}), as well as targeted searches around smaller samples (e.g., \citealt{2016ApJ...823...49S,2018RNAAS...2...41W,2019MNRAS.490.1066D,Brandner2021}). 

To date, only two white dwarfs have been observed to undergo transits from circumstellar debris: WD\,1145+017, found from serendipitous {\em K2} observations \citep{Vanderburg2015} and ZTF\,J0139+5245 \citep{Vanderbosch2020}, found from serendipitous Zwicky Transient Facility (ZTF) photometry. Additionally, analysis of {\em TESS} data revealed transits from a companion straddling the planet/brown-dwarf boundary in a 1.4\,day orbit around WD\,1856+534 \citep{2020Natur.585..363V}.

Finding more white dwarfs with transiting circumstellar planetary debris will provide vital constraints on the dynamics and composition of planets and planetesimals during the final stages of stellar evolution \citep{2016NewAR..71....9F}. We expect that as the host stars begin their evolution into white dwarfs they will engulf any close-in planets out to roughly 1.5\,AU \citep{2012ApJ...761..121M}, so the debris we see that survived this phase likely had its orbit perturbed inward towards the Roche limit of the white dwarf \citep[e.g.,][]{2012ApJ...747..148D}.

In the era of big-data astronomy, developing tools to efficiently mine all-sky surveys to discover more variable white dwarfs without computing periodograms should be prioritized, especially since white dwarfs offer the clearest window into the future and compositions of planetary systems. In this manuscript, we present two such methods that solely use anomalously high levels of scatter in {\em Gaia} DR2 and ZTF DR3 photometry as a proxy for variability.

The first is a global, all-sky proxy for variability that makes use of only {\em Gaia} DR2 photometry, while the second aims to build a more robust variability detection procedure through the inclusion of ZTF photometry and alerts. As a proof of concept, we applied the second method to a sample of $12{,}073$ white dwarfs within 200\,pc and centered on the ZZ Ceti instability strip, and confirm the detection of 19 new ZZ Cetis with follow-up high-speed photometry from McDonald Observatory. Additionally, we report robust evidence for five new white dwarfs that exhibit transit-like dips in their ZTF and/or McDonald light curves. We further discuss the potential for this method to expand the population of both pulsating white dwarfs and white dwarfs harboring transiting debris.

%%%%%%%%%%%%%%%%%%%%%%%%%%%%%%%%%%%%%%%%%%%%%%%%%%%%%%%%%%%%%%%%%%%%%%%%%%%%%%%%%%%

                                %OBSERVATIONS                
                
%%%%%%%%%%%%%%%%%%%%%%%%%%%%%%%%%%%%%%%%%%%%%%%%%%%%%%%%%%%%%%%%%%%%%%%%%%%%%%%%%%%

\section{Observations} \label{sec:observations}

\subsection{Gaia DR2 Photometry}

We selected targets for variability assessment from the {\em Gaia} DR2 catalog of white dwarf candidates \citep{GentileFusillo2019}, which contains $486{,}623$ objects. To ensure an astrometrically clean sample, we applied the quality cuts recommended by \citealt{2018A&A...616A...2L} and  \citealt{2018A&A...616A...4E} (see Appendix~\ref{sec:gaiacuts}) and restricted our sources to those within 200\,pc. This resulted in a list of $46{,}002$ all-sky sources which we use to assess variability using {\em Gaia} DR2 data products alone, with observations spanning 22 months from 2014 July to 2016 May \citep{2018A&A...616A...1G}.

From this 200\,pc sample, we also generated a separate list of sources with photometrically determined effective temperatures near to or within the ZZ Ceti instability strip, where DA white dwarfs are found to pulsate. For this cut, we used the $T_{\mathrm{eff}}$ determinations provided by \citet{GentileFusillo2019} assuming H-atmospheres and restricted the 200\,pc sample to those with $7000 \leq T_{\mathrm{eff}} \leq 16{,}000$. This sample forms the basis of our variability assessments using both ZTF and {\em Gaia} data products, so we also mandated all objects to have $15.0 \leq$ \textsc{phot\_g\_mean\_mag} $\leq 21.0$ and \textsc{dec} $> -25.0$ to stay within ZTF's operational limits \citep{Bellm_2019}. These cuts resulted in an object list with $18{,}269$ sources.

\subsection{Public ZTF Photometry}
For the $18{,}269$ objects defined in our sample centered on the ZZ Ceti instability strip, we queried the public ZTF survey \citep{Masci2019,Bellm_2019,2019PASP..131g8001G} for the DR3 $g$ and $r$-band light curves by performing 3\,arcsecond radii cone searches centered on the {\em Gaia} DR2 \textsc{ra} and \textsc{dec} using the API provided by IRSA.\footnote{\href{https://irsa.ipac.caltech.edu/docs/program_interface/ztf_lightcurve_api.html}{IRSA ZTF Light Curve Queries API}} The coverage for ZTF DR3 extends from 2018 March 17 to 2019 December 31.

To remove any erroneous or potentially contaminated observations, we applied highly conservative filtering to these light curves. For every exposure, we required: $\textit{catflags} = 0$, $|\textit{sharp}| < 0.25$, and $\textit{mag} < \textit{limitmag} - 1.0$. The \textit{catflags} and \textit{sharp} constraints are both recommended cuts for clean light curve extractions in the ZTF Science Data System Explanatory Supplement (ZSDS)\footnote{\url{http://web.ipac.caltech.edu/staff/fmasci/ztf/ztf_pipelines_deliverables.pdf}}. We chose a slightly more restrictive cut on \textit{sharp} than the recommended values of $|\textit{sharp}| < 0.5$ to ensure better removal of cosmic ray contaminated and elongated sources. The constraint on \textit{limitmag} was determined through trial and error after noting several objects which exhibited artificial flux increases when the measured magnitude was within 1.0\,mag of an exposure's \textit{limitmag}.

Subsequently, we separated all of the light curves by filter type and sorted them by object ID (\textit{oid}). Similar to the first \textit{limitmag} constraint, we then discarded any observations where $\textit{limitmag}-1$ was less than the median observed magnitude for a respective \textit{oid} and filter to remove spuriously high flux measurements. To ensure the light curves queried at each {\em Gaia} source location all belonged to the same source, we compared the median ZTF $g$ and $r$ magnitudes for each \textit{oid} to the catalogued Pan-STARRS1 (PS1) DR2 $g$- and $r$-band PSF-fit magnitudes. ZTF magnitudes are calibrated using PS1 photometry and generally agree within 0.10\,mag out to $g \approx 20, r \approx 20$ \citep{Masci2019}. If the median ZTF $g$ and $r$ magnitudes for a given \textit{oid} differed from the respective PS1 magnitudes by more than 0.25-mag, any observations pertaining to such an \textit{oid} were discarded to decontaminate the light curve from observations of neighboring objects. Moreover, only decontaminated light curves with at least 20 observations were considered in this analysis, i.e. $n_{\mathrm{obs},g} \ \mathrm{or} \ n_{\mathrm{obs},r} \geq 20$, to allow meaningful light curve statistics and reduce the chance of including false positive detections in our analysis \citep{2020ApJS..249...18C}.

Finally, we accessed the public ZTF transient alert database \citep{2019PASP..131a8001P} using the API provided by the Las Cumbres Observatory Make Alerts Really Simple (MARS) project\footnote{\url{https://mars.lco.global/}} to query alert packets for each of the $18{,}269$ objects, granting us alert packets triggered from 2018 June 1 to 2020 August 4.

\subsection{Final Object Sample and Decontamination}
\label{sec:ps1_decontamination}
As a final decontamination check, we performed automated 1\,arcminute cone search queries of PS1 photometry \citep{Chambers2016,Flewelling2020} to identify any nearby objects that could contaminate both the {\em Gaia} DR2  and ZTF photometry with artificial excessive scatter, e.g. extremely bright stars or exceptionally close stars. Similar to the ZTF light curve filtering, we developed a system of empirically motivated criteria to remove potentially contaminated objects from our catalog. These criteria are outlined in Appendix~\ref{sec:ps1criteria}.

After this automated decontamination step, we inspected the ZTF light curves and images for every object in our top 1\% highest ranked sample (183 objects, not to be confused with the final top 1\% presented in Section~\ref{sec:combined_results}) by hand to identify any remaining pathologies. We found several objects which had passed the PS1 search criteria in Appendix~\ref{sec:ps1criteria}, but still had exceptionally close stars or nearby bright stars which were not catalogued by PS1. In addition, several objects exhibited large flux dropouts which, upon ZTF image inspection, were caused by bad pixel columns crossing the stellar PSF. Lastly, several objects near CCD edges were contaminated by unmasked ghosts causing excess light curve scatter. Such artifacts are expected due to very bright stars that fall within CCD gaps or beyond the focal-plane edge (see section 6.5 of the ZSDS). 

In the end, the automated decontamination procedure flagged 4258 objects, and an additional 1926 objects lacked a sufficient number of ZTF observations, while the aforementioned inspection steps identified a total of 12 out of 183 contaminated objects in the top 1\%. As a result, our final {\em Gaia}+ZTF sample contains a total of $12{,}073$ objects.

\begin{figure*}[!ht]
    \figurenum{1}
    \centering
    \epsscale{1.15}
    \plotone{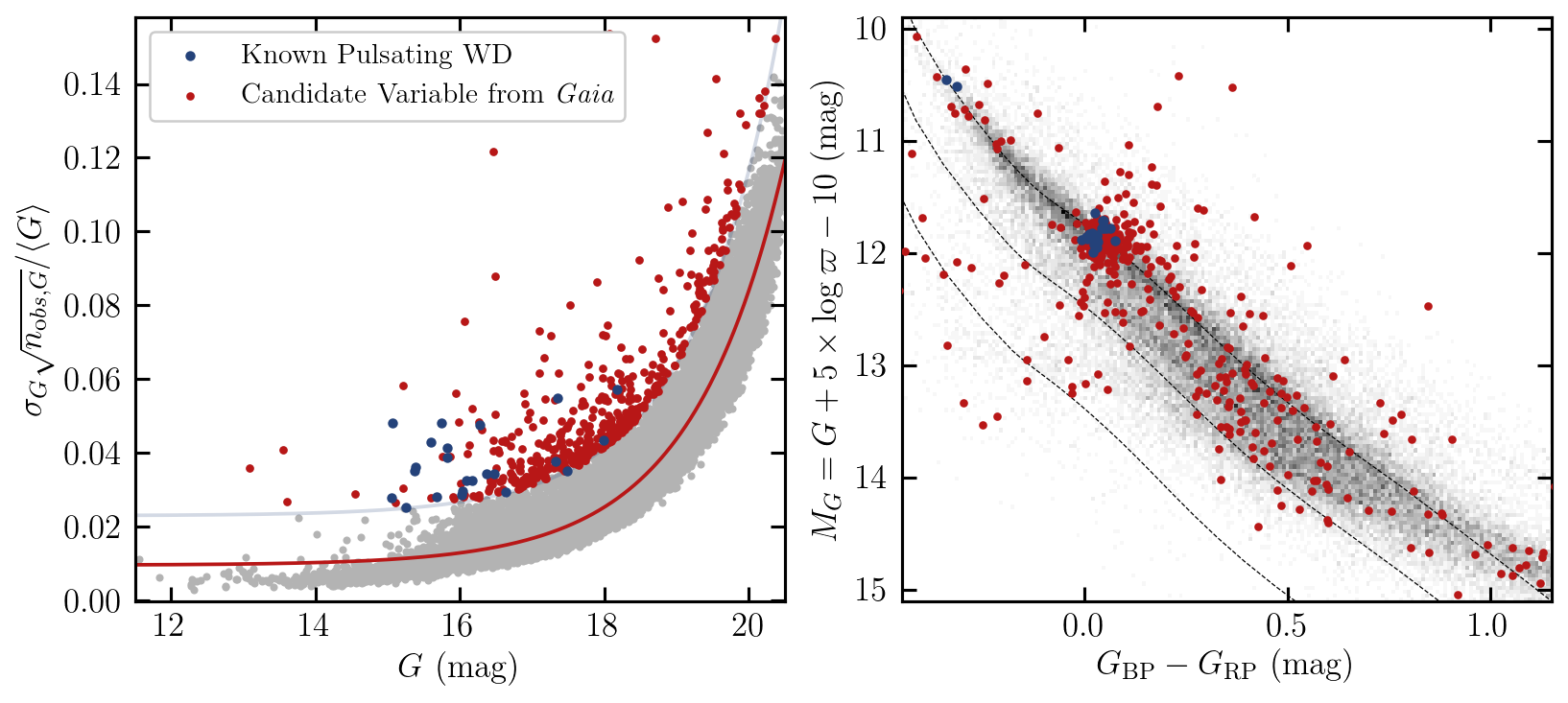}
    \caption{{\bf Left}: Empirical {\em Gaia} DR2 photometric variability metric as a function of magnitude for all white dwarfs in the 200\,pc sample in grey. More details on the exponential fit are described in Appendix~\ref{sec:varindex}. The 1\% most variable white dwarfs are marked in red, and known pulsating white dwarfs in blue. {\bf Right:} The 1\% most variable white dwarfs defined at left are shown in the {\em Gaia} color-magnitude diagram. The dashed lines show hydrogen-atmosphere white dwarf cooling tracks from \citet{2011ApJ...730..128T} showing, from top to bottom, $\log(g)\,=\,8.0,\,8.5,\,9.0$, corresponding to white dwarf masses of roughly 0.6, 0.9, and 1.2~$M_{\odot}$, respectively.}
    \label{fig:gaia_only_variables}
\end{figure*}

\subsection{Follow-up Observations}
Over the course of two years, from October 2018 to November 2020, we observed 34 white dwarfs in this study from McDonald Observatory on the 2.1\,m Otto Struve telescope using the ProEM Camera at Cassegrain focus in order to obtain high-speed, time-series photometry to assess variability. While most observations were conducted using the blue broad-bandpass {\em BG40} filter, we obtained multi-color photometry for some objects using the SDSS $g$-, $r$-, and $i$-band filters. These 34 objects were originally identified as variable white dwarf candidates based on excess {\em Gaia} photometric scatter (Equation~\ref{eq:gaia}) and their triggering of ZTF transient alerts. In particular, objects with $T_{\mathrm{eff}} \simeq 11{,}000$\,K and $\log(g) \simeq 8.0$ were prioritized to see if they pulsated at periodicities characteristic of outbursting ZZ Cetis \citep{2015ApJ...809...14B, 2015ApJ...810L...5H, 2016ApJ...829...82B}.

Following dark and flat-field corrections using standard \textsc{iraf} procedures, we performed aperture photometry over a range of circular apertures using the {\scshape ccd\_hsp iraf} routine \citep{2002A&A...389..896K}. Employing the Python package \textsc{phot2lc} (Vanderbosch et al.~in prep.)\footnote{\url{https://github.com/zvanderbosch/phot2lc}}, the light curves were sigma-clipped to 4$\sigma$ using a moving window of width 25 data points, clipped by-hand of clearly spurious data points, and detrended with a low-order polynomial fit to account for airmass changes. Within \textsc{phot2lc}, we used \textsc{Astropy} \citep{2013A&A...558A..33A, 2018AJ....156..123A} to apply barycentric corrections to the mid-exposure time stamps of each image. The optimal aperture that minimized the average point-to-point scatter was selected for light-curve extraction. Light curves are presented in Appendix~\ref{sec:mcd_lcs} with accompanying periodograms.

An additional eight white dwarfs were identified as variable candidates using the global {\em Gaia} variability metric. We observed these objects using the Small and Moderate Aperture Research Telescope System (SMARTS) consortium 0.9\,m telescope at Cerro Tololo Inter-American Observatory in May 2018. We obtained time-series photometry through a Johnson-$V$ filter using the Tek2K CCD photometer at Cassegrain focus. These data were bias and flat-field corrected using standard \textsc{iraf} routines. Aperture photometry was then conducted using the \textsc{photutils} Python suite \citep{Bradley2020} over a range of circular aperture radii. The light curves were sigma clipped to 4$\sigma$ and the aperture that minimized the average point-to-point scatter was selected for analysis. These light curves and corresponding periodograms are presented in Appendix~\ref{sec:ctio_lcs} along with a table of stellar parameters.

\begin{figure*}[t]
    \figurenum{2}
    \epsscale{1.15}
    \plotone{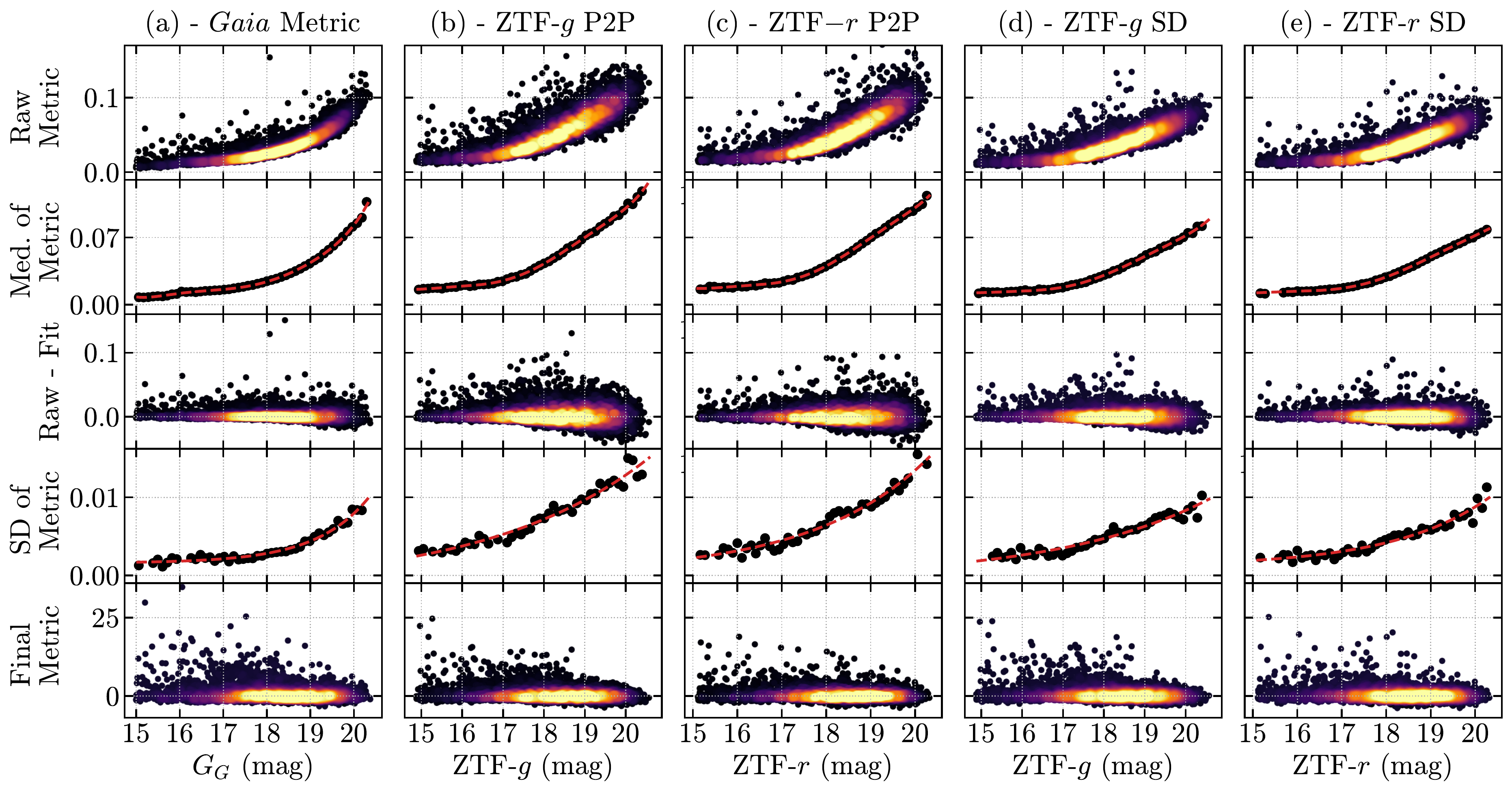}
    \caption{A visualization of the {\em Gaia} and ZTF point-to-point scatter (P2P) and standard deviation (SD) metrics, detailed in Sections~\ref{sec:vgaia} \&~\ref{sec:vztf}. Each column pertains to an individual filter and metric, while each row catalogues the evolution of the metrics with each step in the detrending process, beginning with the raw metrics, then the fitted medians of the binned raw metrics, the differences between the above panels, the fitted standard deviations within each bin, and finally quotients of the above difference and fit. The distributions are color coded such that the brighter points indicate relatively dense regions, while the dark points indicate sparse regions. Note that a small number of objects with exceptionally large metric values are not visible on these y-axis scales.}
    \label{fig:detrending}
\end{figure*}

For some objects exhibiting variability indicative of transiting planetary debris or other non-pulsational phenomena, we obtained identification spectra using the second-generation Low-Resolution Spectrograph (LRS2; \citealt{2016SPIE.9908E..4CC}) on the 10.2-m Hobby-Eberly Telescope (HET) at McDonald Observatory. LRS2 is fed by a microlens-coupled bundle of 280 fibers, each $0.6''$ in diameter, with a unity fill factor over a $6''\times12''$ field of view. We used the blue LRS2 spectrograph (LRS2-B) which provides full coverage over 3700--7000\,\AA\ with two separate arms, UV and orange, overlapping between 4600 and 4700\,\AA. With two-point binning in the UV arm, we achieved spectral resolutions of about 4.4 and 5.1\,\AA\ in the UV and orange arms, respectively. We used LRS2-B to observe the transiting debris candidate ZTF\,J0923+4236 on 2020 November 2 with five consecutive 600-s exposures in $1.7\arcsec$ seeing, and to obtain time resolved spectroscopy of the polar candidate ZTF\,J0146+4914 on 2020 November 7 with 13 consecutive 320-s exposures in $2.1\arcsec$ seeing. The spectra were reduced using the \textsc{Panacea} reduction pipeline (G.~Ziemann et al. 2021, in preparation)\footnote{\url{https://github.com/grzeimann/Panacea}}.

We also obtained follow-up observations using the DeVeny Spectrograph mounted on the 4.3\,m Lowell Discovery Telescope (LDT,  \citealt{2014SPIE.9147E..2NB}). Using a 400 line mm$^{-1}$ grating and a 1\arcsec\ slit, we obtained a roughly 4.6\,\AA\ resolution. Observations were carried out on 2020 November 16 with $1.7\arcsec$ seeing, with $6\times180$\,s exposures of ZTF\,J0328$-$1219 and $9\times300$\,s exposures of ZTF0347$-$1802. Our spectra were debiased and flat-fielded using standard {\sc starlink} routines \citep{2014ASPC..485..391C}, and were optimally extracted \citep{1986PASP...98..609H} using the software {\sc pamela}. Using {\sc molly} \citep{1989PASP..101.1032M} we applied a wavelength and heliocentric correction.

%%%%%%%%%%%%%%%%%%%%%%%%%%%%%%%%%%%%%%%%%%%%%%%%%%%%%%%%%%%%%%%%%%%%%%%%%%%%%%%%%%%

                                %METRICS              
                
%%%%%%%%%%%%%%%%%%%%%%%%%%%%%%%%%%%%%%%%%%%%%%%%%%%%%%%%%%%%%%%%%%%%%%%%%%%%%%%%%%%

\section{Variability Metrics} \label{sec:metrics}

\subsection{{\em Gaia} Variability Metric}\label{sec:vgaia}

Our initial catalog of empirical white dwarf variability was constructed solely from observations released in {\em Gaia} DR2. We identified candidate variable white dwarfs by selecting white dwarfs with anomalously large photometric errors at a given reported mean magnitude \citep{Hermes2018}. Specifically, we employed {\em Gaia}'s broadband $G$ filter photometry \citep{2018A&A...616A...4E}, using the \textsc{phot\_g\_n\_obs}, \textsc{phot\_g\_mean\_flux}, \textsc{phot\_g\_mean\_flux\_error}, \textsc{phot\_g\_mean\_mag} (henceforth $n_{\mathrm{obs},G}, \ \langle G \rangle, \ \sigma_G, \ G$, respectively) entries in particular. By definition, $\sigma_G$ is an empirically determined value: the standard deviation of the $G$-band flux measurements normalized to $\sqrt{n_{\mathrm{obs},G}}$ \footnote{\href{https://gea.esac.esa.int/archive/documentation/GDR2/Gaia_archive/chap_datamodel/sec_dm_main_tables/ssec_dm_gaia_source.html}{{\em Gaia} DR2 Data Release Documentation 14.1.1 \textsc{gaia\_source}}} \citep{2016A&A...595A...7C, 2018A&A...616A...4E}. Therefore, $\sigma_G$ is a quantification of the scatter in the  {\em Gaia} photometry of individual sources (see also \citealt{2020svos.conf...11E,Mowlavi2020,Andrew2021}).

Our {\em Gaia} variability metric is shown in Figure~\ref{fig:gaia_only_variables} and defined as\footnote{ Equation \ref{eq:gaia} is identical to the {\em Gaia} DR2 variability proxy employed by \citet{Mowlavi2020}.}:
\begin{equation}
\label{eq:gaia}
    V_G \equiv \frac{\sigma_G}{\langle G \rangle} \sqrt{n_{\mathrm{obs},G}}
\end{equation} 
The complicated scanning pattern that the {\em Gaia} spacecraft undertakes means some stars have been observed far more times than others, motivating us to normalize by the number of observations ($n_{\mathrm{obs},G}$). The 22 months of photometry reported in {\em Gaia} DR2 has been iteratively clipped to 5$\sigma$ (see Section 5.3.5 of the {\em Gaia} DR2 release documentation, \citealt{2018gdr2.reptE...5B}), so it is possible that the most variable white dwarf objects and systems are absent or underestimated by this metric. Still, objects with anomalously high $\sigma_G$ values relative to similarly bright objects are candidates for excess scatter caused by variability. 

To identify candidate variables, we use a double-exponential function (defined explicitly in Appendix~\ref{sec:varindex}) to remove trends with magnitude and define our {\em Gaia} variability proxy, \textsc{varindex}. The double-exponential, shown in light blue in the left panel of Figure~\ref{fig:gaia_only_variables}, enables us to define the 1\% most variable white dwarfs compared to this baseline. The position of these most-variable objects within the local (200\,pc) sample of {\em Gaia} white dwarfs is shown in the right panel of Figure~\ref{fig:gaia_only_variables}. The vast majority of variable objects are found near the ZZ Ceti (DAV) instability strip, where pulsations are known to cause high-amplitude optical variability, validating our method. Results when using this metric as a global proxy for variability are reported in Section~\ref{sec:gaia_results}. 

\begin{deluxetable*}{cccccccccccc}
\tablenum{1}
\tablecaption{Sample table of parameters and metrics for the top 1\% most variable white dwarfs from our joint {\em Gaia}+ZTF metrics. See Appendix~\ref{sec:top1table} for the full table and description of columns. \label{tab:sampletop1}}
\tablewidth{0pt}
\tablehead{
\colhead{WD} & \colhead{$\alpha$ (deg)} & \colhead{$\delta$ (deg)} & \colhead{Class} & \colhead{$R$} & \colhead{$\tilde{V}_{\mathrm{ZTF}}$} & \colhead{$\tilde{V}_{G}$} & \colhead{$N_{A}$} & \colhead{$G$} & \colhead{$T_{\mathrm{eff}}$} & \colhead{$\log(g)$} & \colhead{Observed}
}
\startdata
 WD\,J001038.25+173907.24    &  2.65952 &  17.65175 & cZZ   &  27.7 &  4.1 &  9.7 &  1 & 17.8 &   11220 &  7.9 & McD \\
 WD\,J002511.11+121712.39    &  6.29599 &  12.28661 & CV    & 139.7 &  9.8 & 25.2 &  3 & 17.5 &    8910 &  7.3 & \\
 WD\,J002535.80+223741.89    &  6.39950 &  22.62813 & V     &  15.2 &  4.1 &  3.5 &  1 & 18.0 &   11490 &  8.1 & \\
 WD\,J004711.37+305609.18    & 11.79746 &  30.93552 & cZZ     &  20.1 &  3.9 &  6.1 &  1 & 17.6 &   10450 &  7.5 & ZTF \\
 WD\,J010207.20$-$003259.57  & 15.53151 &  -0.55041 & ZZ    &  14.6 &  3.3 & 11.3 &  0 & 18.2 &   10320 &  7.9 & \\
\enddata
\end{deluxetable*}

Finally, we have constructed an additional {\em Gaia} DR2 variability metric, $\tilde{V}_G$, extended to our overlapping ZTF sample. Beginning with the raw $V_G$ values obtained from Equation~\ref{eq:gaia}, we built this metric by first subtracting a sixth-order polynomial fit to the median $V_G$ values within 50 iteratively sigma clipped ($\sigma_{\mathrm{upper}} = 3, \sigma_{\mathrm{lower}} = 100$) magnitude bins of equal width using \textsc{lmfit} \citep[][see left panel of Figure~\ref{fig:detrending}]{newville_matthew_2014_11813}. We then divided by an exponential fit to the standard deviation of the values in each bin, yielding the final {\em Gaia} DR2 variability metric, $\tilde{V}_G$, for our {\em Gaia}+ZTF sample.

\subsection{ZTF Light Curve Metric}\label{sec:vztf}
The ZTF light curve metric is based on two measurements of excess scatter: the average point-to-point scatter ($\tilde{V}_{\mathrm{P2P}}$) and the standard deviation in the normalized DR3 light curves ($\tilde{V}_{\mathrm{SD}}$). Ultimately, whichever of the two metrics had the largest value was selected to represent the ZTF light curve metric, $\tilde{V}_{\mathrm{ZTF}}$, for an object: $\tilde{V}_{\mathrm{ZTF}} \equiv \textsc{max}(\tilde{V}_{\mathrm{SD}}, \tilde{V}_{\mathrm{P2P}})$. This was motivated to allow for the detection of both short- and long-term variability, so as not to avoid transient and transiting systems (see Section~\ref{sec:transits}).

We applied an identical detrending routine to both the point-to-point scatter and standard deviation metrics independently in order to remove trends with magnitude. First, we normalized each ZTF $g$ and $r$-band light curve to their respective median magnitudes, resulting in relative flux units, and then proceeded to calculate the raw $g$ and $r$ light curve metrics. The raw metrics were then organized into 50 magnitude bins of equal width (covering $15.11 < r < 20.32$ with an average of 235 objects per bin) still separated by filter. This formed the four ZTF columns in Figure~\ref{fig:detrending}. After iteratively sigma clipping each bin ($\sigma_{\mathrm{upper}} = 3, \ \sigma_{\mathrm{lower}} = 100$), we fit the median metric value of each bin with a sixth-order polynomial as a function of magnitude using the least-squares fitting Python module \textsc{lmfit}. We subtracted this polynomial from the raw metrics to remove the trend in scatter with magnitude. This is visualized in the second and third rows of Figure~\ref{fig:detrending}.

Even with this polynomial subtraction, a trend in the standard deviations of the metrics with magnitude remained. To remove this second trend we fit an exponential curve as a function of magnitude to the standard deviations of the metric values within the same magnitude bins from before using \textsc{lmfit}, as depicted in row 4 of Figure~\ref{fig:detrending}. To ensure a high quality fit, bins with fewer than 5 data points were excluded from this fitting procedure and outliers were removed by hand. The bins excluded in these fits were removed only to improve the fitting process, but the objects within them were not removed from our variability analysis. This prevented bins with poorly defined statistics from influencing the fitting process. We then divided the metric values by the fitted exponential to normalize the metrics onto scales of standard deviation ($\sigma$), yielding the distributions shown in the bottom row of Figure~\ref{fig:detrending}. 

With the detrending complete we combined the respective metrics from the ZTF-$g$ and ZTF-$r$ light curves into a single value through a weighted average normalized to the number of observations in each filter, yielding the final ZTF point-to-point scatter and standard deviation metrics, $\tilde{V}_{\mathrm{P2P}}$ and $\tilde{V}_{\mathrm{SD}}$, respectively. Again, for each object the larger of these two metrics was recorded as its final ZTF light curve metric, $\tilde{V}_{\mathrm{ZTF}}$.

\subsection{ZTF Alert Metric}
The ZTF Alert variability metric, $N_{A}$, is the number of alerts generated by an object that could be attributed to observations of real periodic or transient variability with high confidence. This was accomplished by filtering the alerts using a set of criteria outlined by the ZSDS and IPAC\footnote{\url{http://web.ipac.caltech.edu/staff/fmasci/ztf/ztf_pipelines_deliverables.pdf}}: where $rb \geq 0.65$, $n_{\mathrm{bad}} = 0$, \textit{fhwm} $\leq 5$ pixels, \textit{elong} $ \leq 1.2$, and $|\textit{magdiff}| \leq 0.1$ mag.

Even though all observations that trigger alerts are already filtered by IPAC, these additional criteria construct a sturdy buffer that excludes any potentially false positive observations.

\subsection{Ranking Parameter}
Capitalizing on the fact that $\tilde{V}_G$ and $\tilde{V}_{\mathrm{ZTF}}$ are both set onto scales of $\sigma$, we defined a combined variability metric, $R$, that ranks objects according to their level of variability:
\begin{equation}
    R = (\tilde{V}_G + \tilde{V}_{\mathrm{ZTF}}) \cdot (1 + N_{A})
    \label{eqn:rank}
\end{equation}
Here $N_A$ serves as a boosting factor, rewarding objects that have triggered ZTF transient alerts, while not penalizing objects whose transient or periodic variability may be too low amplitude to trigger alerts, or whose ZTF coverage may have been too sparse for frequent detections of large flux changes. Likewise, by combining these individual metrics we are able to rank variability in objects sampled over years of observations. The results pertaining to this ranking parameter are presented in Section~\ref{sec:combined_results}, and a sample table of parameters from the top 1\% most variable objects are found in Table~\ref{tab:sampletop1}. 

%%%%%%%%%%%%%%%%%%%%%%%%%%%%%%%%%%%%%%%%%%%%%%%%%%%%%%%%%%%%%%%%%%%%%%%%%%%%%%%%%%%

                                %RESULTS                
                
%%%%%%%%%%%%%%%%%%%%%%%%%%%%%%%%%%%%%%%%%%%%%%%%%%%%%%%%%%%%%%%%%%%%%%%%%%%%%%%%%%%

\section{Results} \label{sec:results}

\subsection{Top 1\% Ranked Only by {\em Gaia}}\label{sec:gaia_results}

The exquisite photometry collected by {\em Gaia} and released in DR2 has already revolutionized variability studies across the HR diagram, especially by \citet{2019A&A...623A.110G}, who showed a high concentration of variable white dwarfs ($>$50\%) around the ZZ Ceti instability strip. The ZZ Ceti instability strip is characterized by a narrow range of effective temperatures, since pulsations are only driven when the white dwarf develops a sufficiently deep convection zone.

However, reproducing the figures of \citet{2019A&A...623A.110G} poses a challenge, since most parameters (especially the $G$-band inter-quartile range) were not publicly released in {\em Gaia} DR2. A small handful of pulsating white dwarfs are flagged as ``short-timescale variable'' in {\em Gaia} DR2, but since the short-timescale ($<$1\,day) variability processing was oriented towards sources with established periodic variability \citep{Holl2018}, there are rarely enough epochs to classify many variable or pulsating white dwarfs.

We show in Figure~\ref{fig:gaia_only_variables} a remarkable consistency with \citet{2019A&A...623A.110G} using our \textsc{varindex} and {\em Gaia} variability metrics, especially the regions of the {\em Gaia} color-magnitude diagram with the most-variable white dwarfs. The biggest clustering of variables in both cases occurs near the colors\footnote{\href{http://www.astro.umontreal.ca/~bergeron/CoolingModels/}{http://www.astro.umontreal.ca/$\sim$bergeron/CoolingModels/}} of the ZZ Ceti instability strip, near $G_{\rm BP}~-~G_{\rm RP}~=~0.0$\,mag and $M_G~=~12.0$\,mag.

\begin{figure*}[t]
    \figurenum{3}
    \epsscale{1.15}
    \plotone{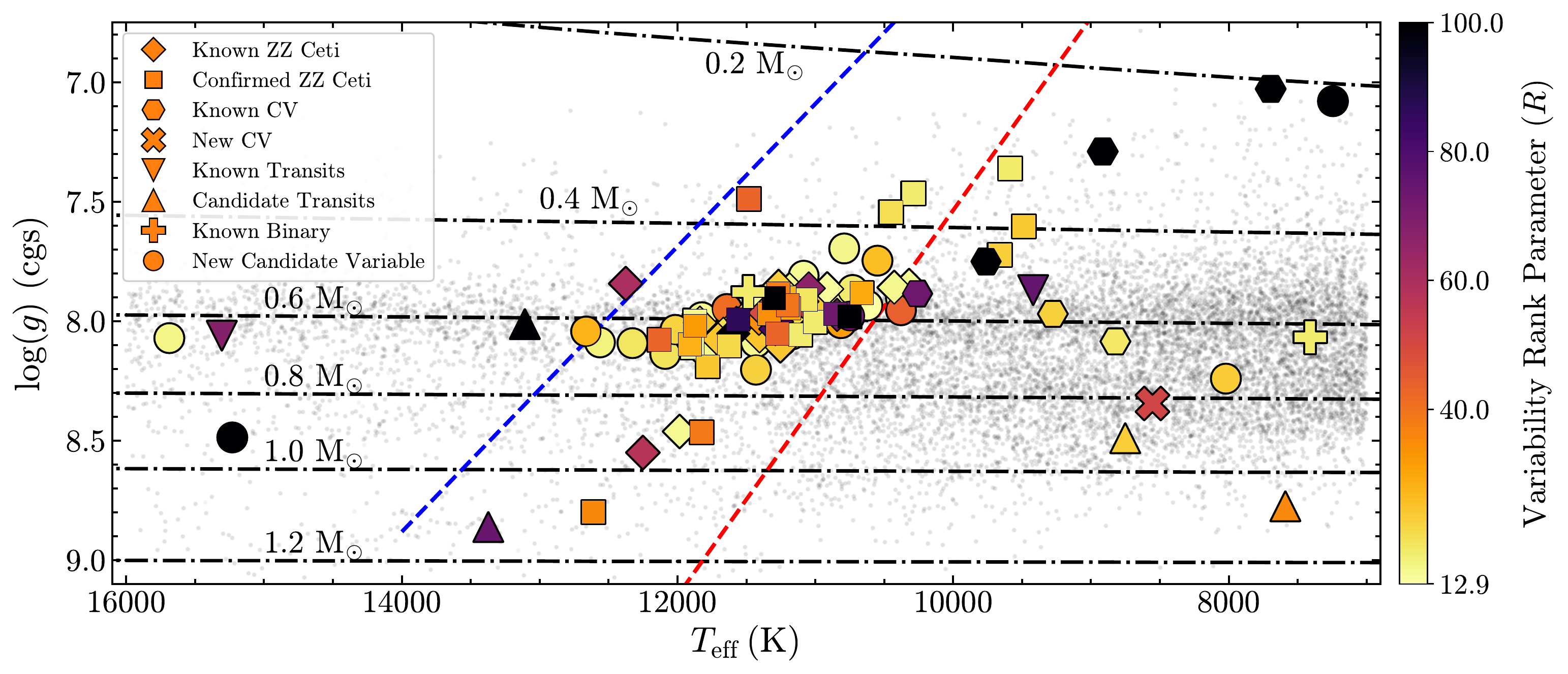}
     \caption{The photometrically determined temperature and surface gravity of our sample, with the top 1\% color-coded by their variability rank parameter value, $R$. The red and blue dashed lines indicate the empirical cool and hot boundaries of the ZZ Ceti instability strip \citep{2015ApJ...812..167G}, while the dashed-dotted black horizontal lines trace theoretical evolutionary cooling tracks for various masses \citep{2011ApJ...730..128T}. The vast majority of our highest-ranked objects are concentrated within the ZZ Ceti instability strip, but numerous non-pulsating variables are found at different temperatures, especially new candidate and known white dwarfs hosting transiting planetary debris. The remainder of our sample is faintly plotted as black circles in the background.}
     \label{fig:loggteff}
\end{figure*}

Nearly one-quarter (23) of the 99 known pulsating white dwarfs within 200\,pc \citep{2016IBVS.6184....1B} meet our criterion for the 1\% most variable white dwarfs by having a \textsc{varindex} $>0.0074$ (see Appendix~\ref{sec:varindex}). We have followed up a number of white dwarfs without previous mention in the literature which have the largest {\em Gaia} \textsc{varindex}, using high-speed photometry, especially from CTIO in the southern hemisphere, and confirm eight new pulsating white dwarfs. All are detailed in Appendix~\ref{sec:mcd_lcs}.

Our double-exponential calibration should define any object with \textsc{varindex} $>0.0$ as a strong candidate for variability: this value is met by 1423 (3\%) of white dwarfs within 200\,pc. This fraction is similar to the results of \citet{2017MNRAS.468.1946H}, who show from {\em Kepler} and {\em K2} observations that ($>97$\%) of non-pulsating and apparently isolated white dwarfs are photometrically constant to better than $1\%$ in the {\em Kepler} bandpass on 1\,hr to 10\,day timescales. Nearly 40\% (38) of the previously known pulsating white dwarfs within 200\,pc have \textsc{varindex} $>0.0$.

Our full implementation of the \textsc{varindex} is likely valid to select variables for any objects in the magnitude range $13\,<\,G\,<\,20$\,mag, and is described in full in Appendix~\ref{sec:varindex}. We also report a calibration of this metric to {\em Gaia} Early Data Release 3, $\textsc{varindex}_{\mathrm{eDR3}}$, in Appendix~\ref{sec:varindex}. Further sections outline our attempts to improve variable white dwarf selection by pairing {\em Gaia} DR2 with ZTF DR3 empirical variability.

\subsection{Top 1\% Ranked by Gaia+ZTF}\label{sec:combined_results}
Out of the ensemble of roughly 12${,}$100 objects with reliable {\em Gaia} and ZTF photometry, we only explored the top 1\% most variable objects in this study, objects with $R \geq 12.9$, as these objects can be assigned high confidence of being variable. Reassuringly, a literature search using SIMBAD revealed 41 of these 121 objects to be known variable white dwarf systems, including ZZ Cetis, cataclysmic variables, eclipsing binaries, and the two known transiting debris systems. These known variables are distributed evenly throughout the sample, with rankings as high as 3 and as low as 118.

\begin{figure*}[ht]
    \figurenum{4}
    \epsscale{1.15}
    \centering
    \plotone{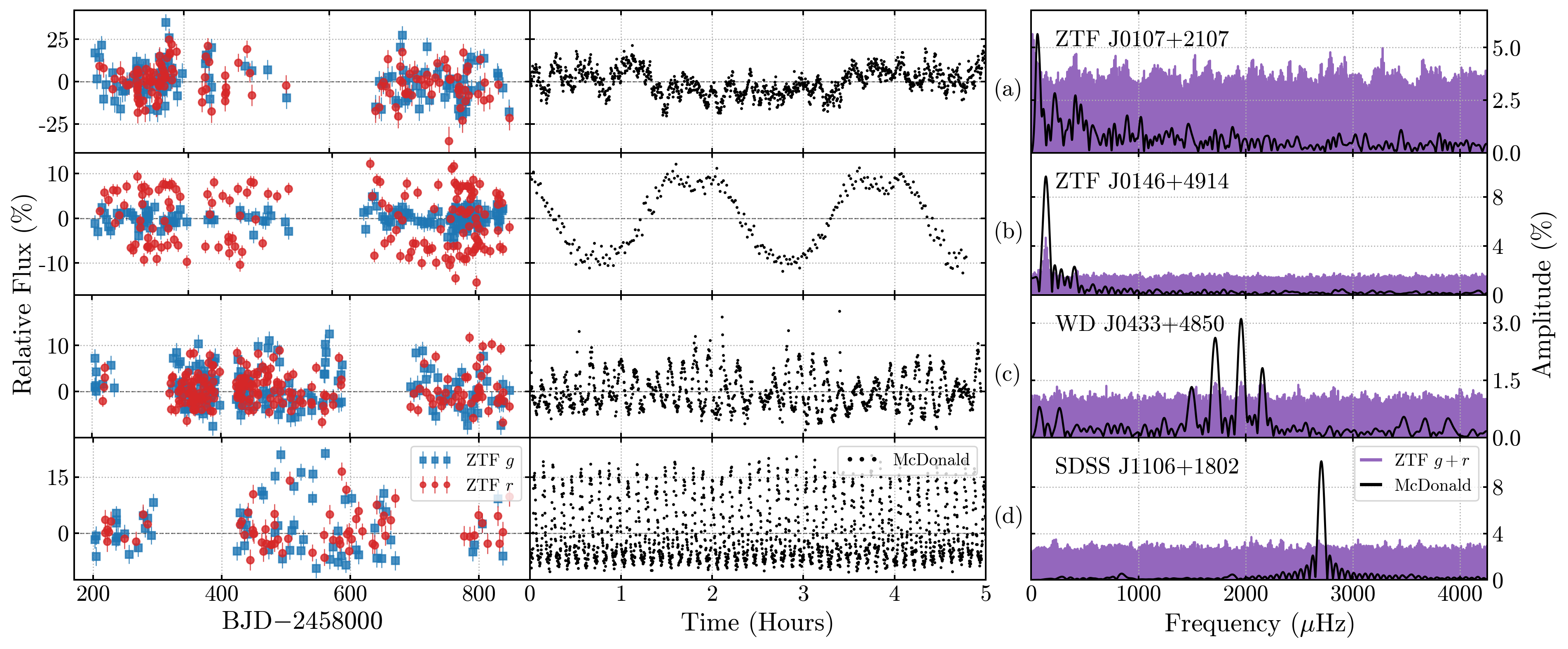}
    \caption{Four examples of new variable white dwarfs, selected from objects in our top 1\% ranked list using {\em Gaia}+ZTF empirical variability. The left-most panel shows ZTF $g$-band (blue squares) and $r$-band (red circles) photometry, and the middle panel shows high-speed, follow-up photometry from McDonald Observatory. The purple periodograms at right represent the combined ZTF $g$- and $r$-band photometry, while the black periodograms represent the McDonald photometry. The four objects showcased here are {\bf (a)} ZTF\,J0107+2107 --- a new transiting debris candidate, {\bf (b)} ZTF\,J0146+4914 --- a new candidate polar, {\bf (c)} WD\,J0433+4850 --- a known ZZ Ceti \citep{Vincent2020}, and {\bf (d)} SDSS\,J1106+1802 --- a new high-mass ZZ Ceti. Notably, even with barycentric corrections applied we do not observe significant pulsation modes in the ZTF periodograms of these two ZZ Cetis. So, our method is able to identify pulsating ZZ Cetis that peak-finding algorithms would otherwise neglect, while also detecting more exotic variability types like those shown in the top two panels.
    \label{fig:example_lcs}}
\end{figure*}

The remaining 86 objects were either unstudied objects or not previously known to be variable. We obtained follow-up, high-speed photometry from the 2.1\,m Otto Struve telescope at McDonald Observatory for 33 of these objects found in this top 1\% subset, and all revealed variability, mostly indicative of ZZ Ceti pulsations. Six of these objects were recently confirmed to be ZZ Cetis \citep{Vincent2020}, further supporting our selection criteria. And while two of these 33 objects do not demonstrate short-term variability, they clearly demonstrate long-term variability in their ZTF photometry. All 121 objects are presented in the table in Appendix~\ref{sec:top1table}, while a sample of this information is presented in Table~\ref{tab:sampletop1}.

\begin{deluxetable}{cc}[b]
\tablecaption{Classification of the top 1\% most variable objects from the {\em Gaia}+ZTF metrics
\label{tab:top1type}}
\tablenum{2}
\tablewidth{0pt}
\tablehead{ \colhead{Object Type} & \colhead{Number} }
\startdata
        ZZ Ceti & 31 known, 29 new confirmed\\
		Cataclysmic Variable & 6 known, 1 new candidate\\
		Eclipsing Binary & 2 known\\
		Transiting Debris & 2 known, 4$^*$ new candidates\\
		Magnetic Spot & 1 new candidate\\
		Unconfirmed Variable & 45 new candidates
\enddata
\tablenotetext{*}{An additional object ranked in the top 1\% from our {\em Gaia}-only method also  appears to have transit-like dips on short timescales in its follow-up McDonald 2.1\,m photometry (Section~\ref{sec:transits}).}
\end{deluxetable}

A categorical breakdown of the top 1\% is shown in Table~\ref{tab:top1type} and further reflected in Figure~\ref{fig:loggteff}. Most of these objects reside inside the ZZ Ceti instability strip, of which the overwhelming majority are known and candidate ZZ Cetis. Outside of the empirical boundaries of the instability strip are mostly non-pulsating objects, such as known and candidate cataclysmic variables, eclipsing binaries, and transiting planetary debris systems. We specifically review our classifications of new candidate transiting debris systems and ZZ Cetis in Sections~\ref{sec:transits} and~\ref{sec:zzcetis}, respectively.

Figure~\ref{fig:example_lcs} further captures the diversity of the top 1\%, showcasing two white dwarfs clearly exhibiting variability not indicative of pulsations --- (a) ZTF\,J0107+2107 and (b)  ZTF\,J0146+4914 --- and two pulsating ZZ Cetis, (c) WD\,J0433+4850, a known ZZ Ceti \citep{Vincent2020}, and (d)  SDSS\,J1106+1802. This figure shows the ability of our method to translate objects with excess scatter in their ZTF photometry in the left panels into real detections of variability in the follow-up time-series observations. As best demonstrated by the bottom two objects, by relying on excess photometric scatter instead of searching for significant periodogram peaks, we are able to identify both non-periodic and short period variable white dwarfs that would otherwise go undetected at the typical sampling rates of {\em Gaia} and ZTF. In doing so, our method is sensitive to variability on timescales ranging from minutes to days.

While our variability ranking metric does not include an assessment of the periodograms of ZTF light curves, we still performed a search for significant periodicities within the ZTF data of the top 1\% to aid in the classification of candidate variables lacking ground-based follow-up. We first applied barycentric corrections to the ZTF time stamps before proceeding to compute the Lomb-Scargle periodogram of the $g$, $r$, and combined $g$+$r$ light curves. We then searched for any signals greater than four times the mean amplitude ($4 \langle A \rangle$) of the respective periodograms. For this exercise all periodograms were computed over a grid of frequencies ranging from $1\,-\,5000\,\mu$Hz, oversampled by a factor of twice the light curve baseline. We also excluded frequencies equivalent to 0.95--1.05 and 0.48--0.52 cycles per day to reject aliases of the typical diurnal sampling rate \citep{Coughlin2020}. This returned 81 out of 121 objects showing periodic variability beyond the $4 \langle A \rangle$ significance threshold.

Considering the elevated noise from the long baselines and typical long-cadence mean sampling rates of the ZTF time-series, we then computed more conservative 0.1\% false alarm probability significance thresholds for each of these 81 objects. To estimate these significance thresholds we used a bootstrap method \citep[see][]{VanderPlas2018,Bell2019}, where the combined $g$+$r$ flux measurements were randomly resampled 10${,}$000 times with replacement using \textsc{Astropy}. We then computed the maximum amplitude values of the Lomb-Scargle periodogram for each bootstrapped light curve keeping the original time stamps for each iteration. The 99.9th percentile of the distribution of maximum amplitudes is then assumed as the 0.1\% false alarm probability level. The mean of the 81 computed significance thresholds was found to be a factor of $4.94 \pm 0.23$ times the original periodogram amplitudes, justifying adopting a significance threshold of $5 \langle A \rangle$ in future analyses to avoid such a computationally expensive procedure.

A final search for significant peaks was then performed on the original light curves using these computed significance thresholds. So as not to count artificial signals from the wide spectral window, we impose a $\pm\,50\,\mu$Hz buffer around each peak. This buffer was widened to $100\,\mu$Hz for objects densely observed at short cadences during at least one night to account for the exaggerated spectral window effects. An automated recursive prewhitening routine was performed to optimize the selection of significant frequencies using least-squares fitting and minimization of summed sinusoids with \textsc{lmfit}. To remove combination frequencies and harmonics, we identified any frequencies that agreed to within the respective 50\,$\mu$Hz or 100\,$\mu$Hz window of a sum, difference, or integer multiple between any signals of greater amplitude. As a result, we report 26 objects in the top 1\% exhibiting significant periods in either their $g$, $r$, or $g$+$r$ periodograms above their respective thresholds. These periodicities are tabulated in Appendix~\ref{sec:pulsation_spectra}.

Since most of these periodicities we report are driven by pulsations, it should be noted that the pulsation modes in ZZ Cetis are commonly observed to vary significantly in frequency and amplitude over long-baseline observations, prompting large uncertainties from the modulations caused by this signal incoherence \citep[see][]{Greiss2014,Hermes2017b}. This could help explain why we do not observe the pulsation modes of object's like WD\,J0433+4850 with the ZTF time-series in Figure~\ref{fig:example_lcs}. However, for SDSS\,J1106+1802 and the shortest period pulsators, it is more likely that the ZTF public survey's typical mean sampling rate of $\approx 3$\,day is is too long to resolve their pulsations even with hundreds of observations.

Nonetheless, this search led us to the confirmation of periodic variability in several white dwarfs we did not follow up, among which is WD\,J062555.04$-$141442.31 (ZTF\,J0625$-$1414), a previously uncatalogued object that demonstrates behavior not indicative of ZZ Ceti pulsations when folded on its most significant periodicity of 4.46\,hr. The remaining candidates are believed to be ZZ Cetis given their computed periodicites and inferred placement in the {\em Gaia} CMD, and are discussed n Section~\ref{sec:zzcetis}.

Our follow-up observations at McDonald Observatory indicated the discovery of six new objects that have temperatures far from the ZZ Ceti instability strip that also demonstrate variability not indicative of pulsations. These observations are shown in Appendix~\ref{sec:mcd_lcs} under the panel dedicated to non-ZZ Cetis. Because our follow-up is limited, however, we could not definitively constrain the detected variability for one object with our photometry alone. WD\,J053432.93+770757.40 (ZTF\,J0534+7707), a new non-pulsating white dwarf, exhibits 0.72\,hr variability that appears to resemble the rotation of a dark spot on the surface of a magnetic white dwarf, e.g. SDSS J1529+2928 \citep{Kilic2015}. Follow-up spectroscopy to search for Zeeman splitting could confirm this classification, but is outside the scope of our work here.

Another non-pulsating white dwarf we discovered is WD\,J014635.73+491443.10 (ZTF\,J0146+4914), which exhibits peak-to-peak variations of ${\approx}\,20\%$ in the $r$-band in its ZTF and McDonald photometry (Figure~\ref{fig:example_lcs}), while its ZTF $g$-band photometry shows no evidence for variability. Only when analyzing the multi-color photometry obtained using the McDonald 2.1\,m telescope through the SDSS-$g$/$r$ filters, phase folded on a period of 2.057\,hr, do more subtle variations of 2.5\% amplitude emerge in the $g$-band.

\begin{figure}[!t]
    \figurenum{5}
    \epsscale{1.15}
    \plotone{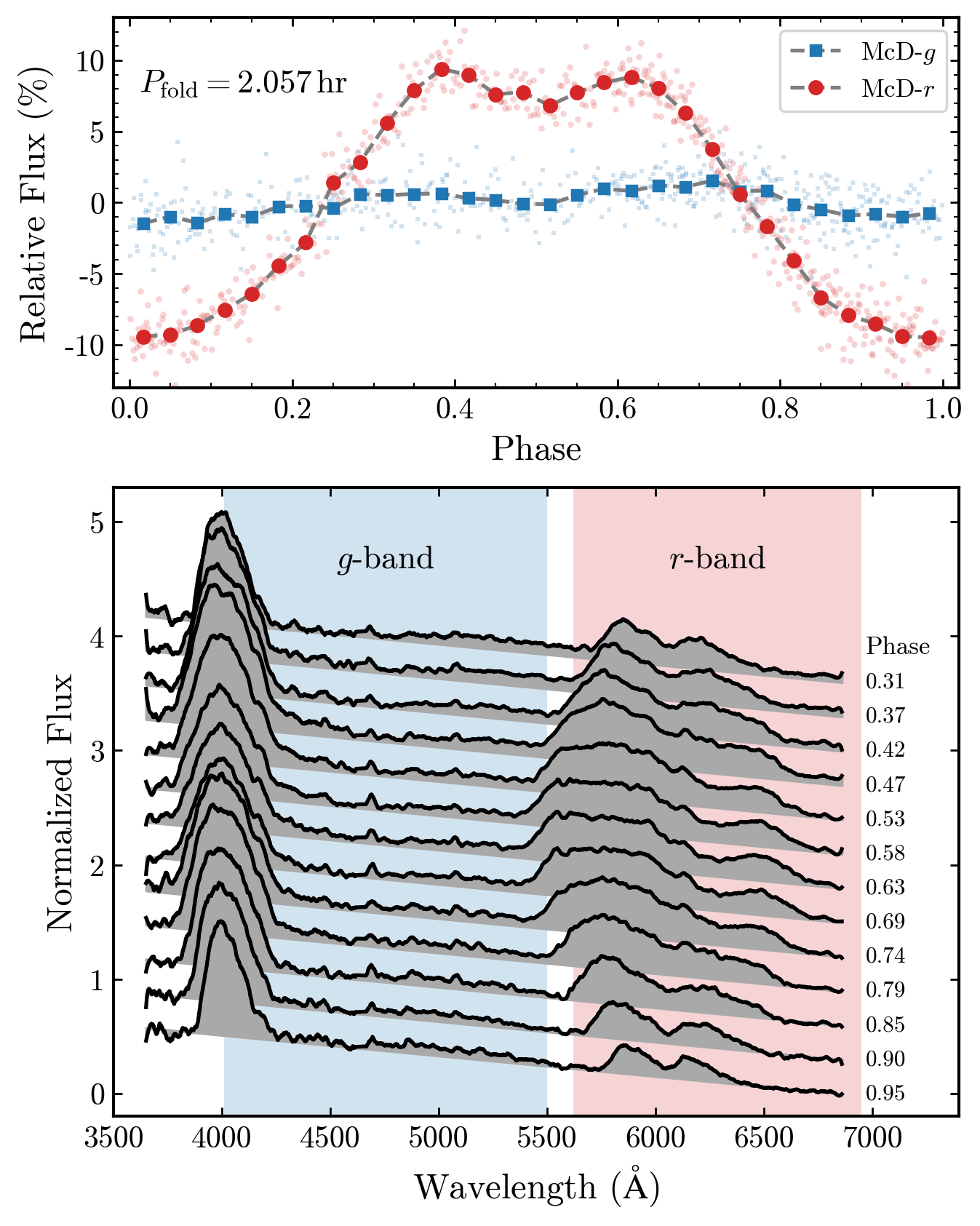}
     \caption{{\bf Top:} Phase-folded McDonald 2.1\,m $g$- and $r$-band photometry of a new polar found from our variability metric, ZTF\,J0146+4914, folded at a period of 2.057\,hr. {\bf Bottom:} Time resolved HET LRS2-B spectroscopy of ZTF\,J0146+4914, each labeled by the phase at mid-exposure, showing cyclotron emission features varying in strength with phase. The features at $\approx$\,6000\,\AA\ and $\approx$\,4000\,\AA\ could correspond to the second and third harmonics, respectively, of cyclotron emission due to a magnetic field of $B~{\approx}~89$\,MG. Colored regions show the extents of the $g$ and $r$ filters, while grey shading under each spectrum is used to accentuate the changing strength of the cyclotron emission.}\label{fig:j0146_folded}
\end{figure}

We show in Figure~\ref{fig:j0146_folded} time-resolved LRS2-B spectroscopy, with each spectrum labeled by the phase at mid-exposure. The strong cyclotron emission features which vary in strength with phase suggest this object is a highly magnetic polar. The two features (at $\approx$\,6000\,\AA\ and $\approx$\,4000\,\AA) could correspond to the second and third harmonics, respectively, of cyclotron emission due to a magnetic field of $B~\approx~89$\,MG \citep{1996MNRAS.282..218F}. Assuming the 2.057\,hr signal is the orbital period, this object falls slightly below the orbital period gap observed for cataclysmic variables (CVs) and lies in the middle of the distribution for magnetic CV systems \citep{Witham2006}. Furthermore, its distance of $56.3\pm0.3$\,pc \citep{Bailer-Jones2018} would potentially establish ZTF\,J0146+4914 as the closest known polar to date \citep{Pala2020,Belloni2020}.

Inspecting the magnitude distributions of our top 1\% sample, we find our {\em Gaia}+ZTF method appears robust down to roughly $G<18.5$\,mag, whereas the {\em Gaia}-only method is sufficient to $G<20.0$\,mag. Our sensitivity to variability, defined as the relative fraction of objects in the top 1\% at a given magnitude, decays quickly when we go fainter than $G\approx18.5$. Beyond this limit we detect two known variable white dwarfs, WD\,J153615.98-083907.53, a CV with $G=18.9$\,mag, and WD\,J013906.17+524536.89 (ZTF\,J0139+5245), one of the two known white dwarfs with transiting planetary debris with $G=18.5$\,mag. These fainter objects exhibit very high amplitude variability, exceeding 30\% in peak-to-peak amplitude. It reasons this is then a requirement for objects fainter than $G\approx18.5$ to be detected by our {\em Gaia}-ZTF method, especially since the associated random noise in the flux measurements of an object grows with magnitude. For the same reason, these objects must manifest larger brightness variations to trigger ZTF alerts, which is also responsible for the dearth of candidates at fainter magnitudes relative to the {\em Gaia}-only technique.

\begin{figure*}[ht]
    \centering
    \figurenum{6}
    \epsscale{1.15}
    \plotone{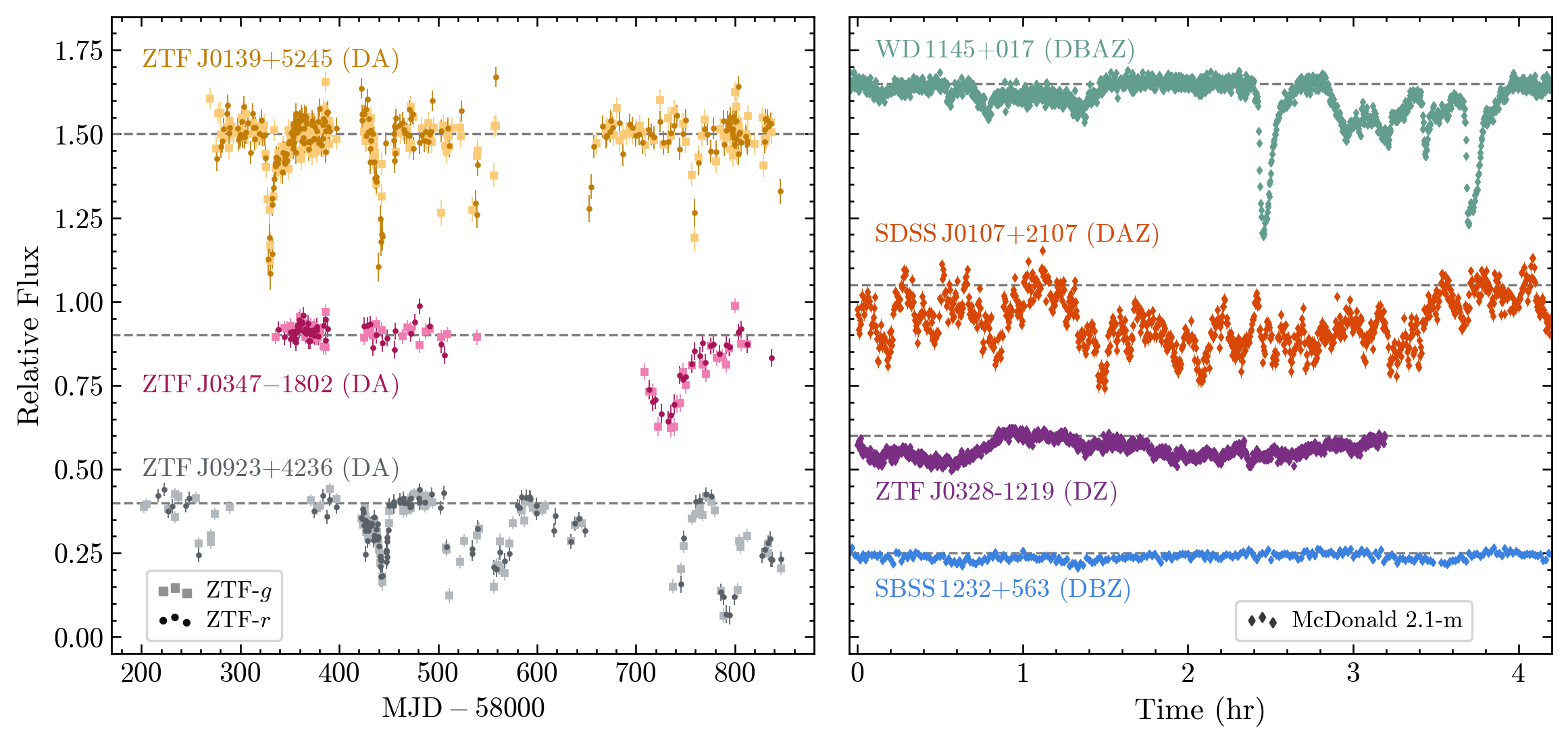}
    \caption{\textbf{Left}: ZTF DR3 light curves for the second known transiting debris host, ZTF\,J0139+5245 (top), along with two transit candidates exhibiting irregularly shaped flux dips on days-long timescales (ZTF\,J0347$-$1802 and ZTF\,J0923+4236). The light curves have been vertically shifted for clarity. Spectral types are shown in parentheses next to each object's name, with references in Table~\ref{tab:transit_params}. \textbf{Right}: McDonald 2.1\,m high speed photometry for the first known transiting debris host, WD\,1145+017 (top), along with three more transit candidates (SDSS\,J0107+2107, ZTF\,J0328$-$1219, and SBSS\,1232+563) which exhibit flux dips on much shorter timescales from minutes to hours. None of these objects are cataclysmic variables, as none show spectroscopic evidence of high accretion rates via emission features (see also Figure~\ref{fig:spectra}).}
    \label{fig:transits}
\end{figure*}

%%%%%%%%%%%%%%%%%%%%%%%%%%%%%%%%%%%%%%%%%%%%%%%%%%%%%%%%%%%%%%%%%%%%%%%%%%%%%%%%%%%

                           % Transits            
                
%%%%%%%%%%%%%%%%%%%%%%%%%%%%%%%%%%%%%%%%%%%%%%%%%%%%%%%%%%%%%%%%%%%%%%%%%%%%%%%%%%%

\section{White Dwarfs with Transiting Debris}\label{sec:transits}
Among the top 1\% most variable {\em Gaia}-ZTF white dwarfs are WD\,J114833.63+012859.42 (WD\,1145+017, \citealt{Vanderburg2015}) and WD\,J013906.17+524536.89 (ZTF\,J0139+5245, \citealt{Vanderbosch2020}), the only two previously known white dwarfs with transiting planetary debris. Accompanying these detections are four additional objects within the top 1\% that exhibit transit-like dips in their ZTF light curves or McDonald 2.1\,m follow-up photometry, shown in Figure~\ref{fig:transits}. An additional object ranked in the top 1\% from our {\em Gaia}-only method also appears to have transit-like dips on short timescales in its McDonald 2.1\,m photometry, and is known to show metal pollution in previous spectroscopic studies. If all five candidates are confirmed, this would more than triple the number of known white dwarfs hosting transiting planetary debris, while increasing the diversity of observed transit durations and recurrence timescales. The light curves of these seven objects are displayed in Figure~\ref{fig:transits}, organized by transit timescales.

\begin{deluxetable*}{ccccccccc}
% \tablenum{2}
\tablecaption{Table of parameters for the known and new candidate white dwarfs with transiting planetary debris. All parameters are sourced from the \citet{GentileFusillo2019} catalog unless otherwise specified below with $\alpha$ and $\delta$ being from the J2015.5 epoch. \label{tab:transit_params}}
\tablewidth{0pt}
\tablenum{3}
\tablehead{
\colhead{Object} & \colhead{$\alpha$ (deg)} & \colhead{$\delta$ (deg)} & \colhead{$d$ (pc)$^{[1]}$} & \colhead{$G$} & \colhead{$T_{\mathrm{eff}}$ (K)} & \colhead{$M_{*,\mathrm{phot}} \ (M_\odot)$} & \colhead{$M_{*,\mathrm{spec}} \ (M_\odot)$} &
\colhead{Spectral Type}
}
\tablecolumns{9}
\startdata
\cutinheadnew{Known White Dwarfs with Transiting Debris}
ZTF\,J0139+5245 & 24.77633 & 52.76027 & 172.9$^{+7.7}_{-7.2}$ & 18.5 & 9420 $\pm$ 580 & 0.52 $\pm$ 0.11 & 0.52 $\pm$ 0.03$^{[2]}$ & DA$^{[2]}$ \\
WD\,1145+017 & 177.13994 & 1.48316 & 141.2$^{+2.5}_{-2.5}$ & 17.2 & $15{,}080$ $\pm$ 640 & 0.64 $\pm$ 0.06 & $\dagger$ & DBZ$^{[3]}$ \\
\hline
\cutinheadnew{New Candidate White Dwarfs with Transiting Debris}
SDSS\,J0107+2107 & 16.95550 & 21.12910 & 90.2$^{+3.9}_{-3.5}$ & 19.2 & 7590 $\pm$ 800 & 1.08 $\pm$ 0.13 & 0.43 $\pm$ 0.03$^{[4]}$ & DAZ$^{[5]}$ \\
ZTF\,J0328$-$1219 & 52.14013 & $-$12.32930 & 43.3$^{+0.2}_{-0.2}$ & 16.6 & 8550 $\pm$ 160 & 0.86 $\pm$ 0.03 & - & DZ$^{[6]}$ \\
ZTF\,J0347$-$1802 & 56.76399 & $-$18.04825 & 76.4$^{+0.8}_{-0.7}$ & 17.4 & $13{,}370$ $\pm$ 510 & 1.13 $\pm$ 0.02 & - & DA$^{[6]}$ \\
ZTF\,J0923+4236 & 140.79749 & 42.60934 & 147.2$^{+2.8}_{-2.7}$ & 17.5 & $13{,}110$ $\pm$ 420 & 0.62 $\pm$ 0.03 & - & DA$^{[6]}$\\
SBSS\,1232+563 & 188.63559 & 56.11195 & 173.0$^{+3.7}_{-3.5}$ & 18.1 & $11{,}670$ $\pm$ 600 & 0.58 $\pm$ 0.07 & $\dagger$ & DBAZ $^{[7]}$
\enddata
\tablecomments{[1] \citet{Bailer-Jones2018}, [2] \citet{Vanderbosch2020}, [3] \citet{Vanderburg2015}, [4] \citet{2019MNRAS.486.2169K}, [5] \citet{2015MNRAS.446.4078K}, [6] this work (see Figure~\ref{fig:spectra}), [7] \citet{2013ApJS..204....5K}, $\dagger$ --- For these objects, spectroscopic $\log(g)$ and mass estimates are highly uncertain due to modeling difficulties of helium absorption lines with metal pollution.}
\end{deluxetable*}

\begin{figure}[ht]
    \centering
    \figurenum{7}
    \epsscale{1.15}
    \plotone{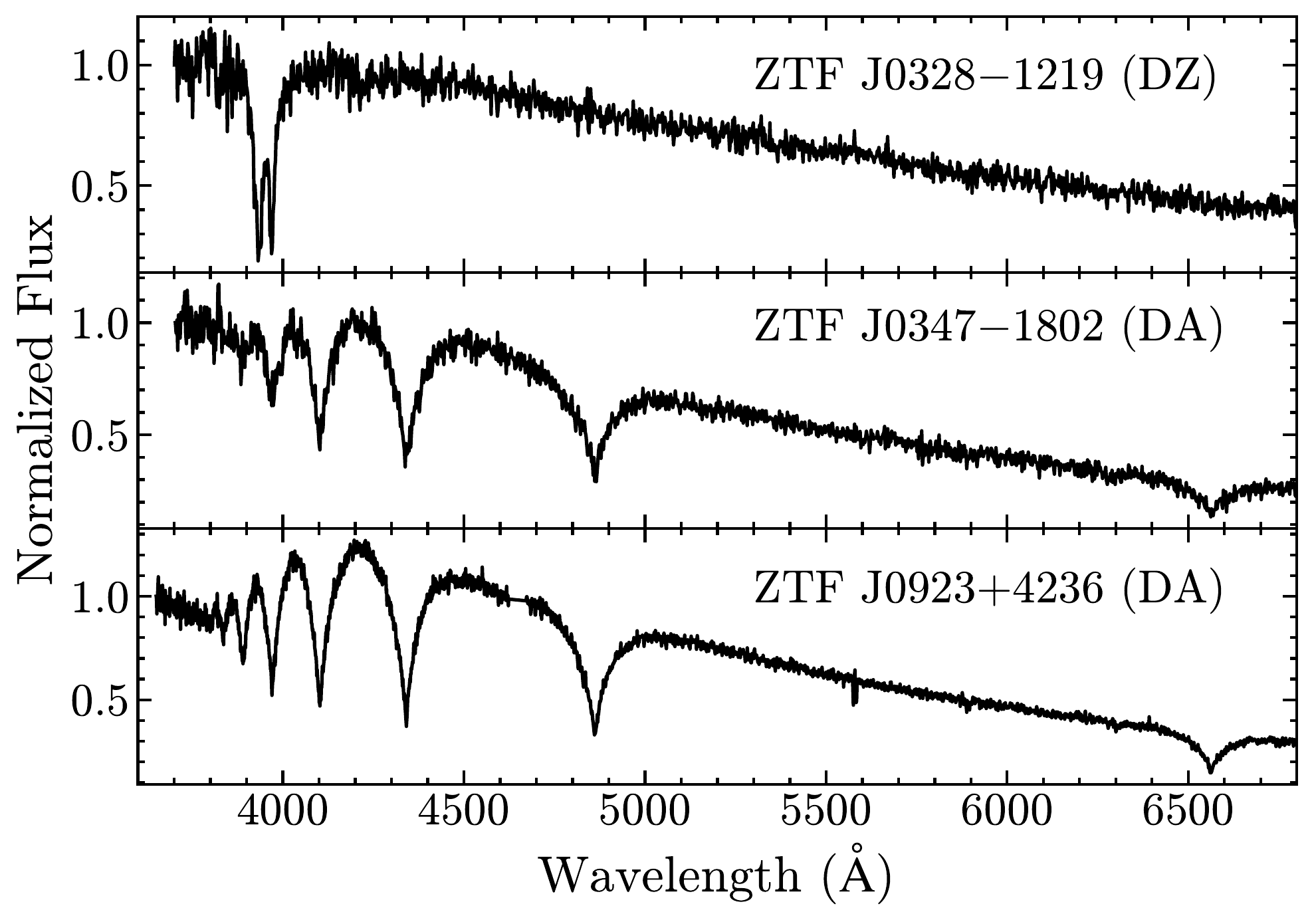}
    \caption{Follow-up identification spectra from LDT and HET of the new transiting debris systems without previous spectroscopy: ZTF\,J0328$-$1219 (top, DZ), ZTF\,J0347$-$1802 (middle, DA), and ZTF\,J0923+4236 (bottom, DA). While ZTF\,J0328$-$1219 is the only to show strong evidence of metal pollution via the prominent Ca\,{\sc ii}~H \& K lines, additional spectroscopic follow-up at higher signal-to-noise is warranted for ZTF\,J0347-1802 and ZTF\,J0923+4236, as the strength of Ca features may be correlated with transit depth if due to circumstellar absorption \citep{Vanderbosch2020}.}
    \label{fig:spectra}
\end{figure}

{\bf ZTF\,J092311.41+423634.16 (ZTF\,J0923+4236)}, the fifth most-variable white dwarf within 200\,pc according to our {\em Gaia}-ZTF ranking parameter, demonstrates multiple transit-like phenomena of varying durations and depths. The highest resolved of these features at MJD $\approx$ 58450 closely resembles a mirror image of the dips observed for ZTF\,J0139+5245 \citep{Vanderbosch2020}: this object displays a gradual ingress and sharp egress, also contrasting the transit shapes of WD\,1145+017 (Figure~\ref{fig:transits}) and the K dwarf KIC\,12557548 \citep{Rappaport2012,vanLieshout2016}. Unlike ZTF\,J0139+5245, the observed transits in ZTF\,J0923+4236 appear more incongruous and complex, repeating on irregular intervals with varying shapes. The dip at MJD ${\approx}\,58450$ appears to have a secondary feature before the onset of the larger dip. This more complicated structure, which appears to be transiting on the order of days, implies ZTF\,J0923+4236 may be at an evolutionary stage in between WD\,1145+017 (roughly 4.5-hr orbital period) and ZTF\,J0139+5245 (roughly 107-day orbital period). We followed up ZTF\,J0923+4236 with high-speed photometry on $\mathrm{MJD}\,{=}\,59138$ and $59139$ using the McDonald 2.1\,m telescope to assess short-term variability, observing small amplitude variations on timescales of $\approx$1 hour (Appendix~\ref{sec:mcd_lcs}). These observations, however, were constrained to less than two hours in length due to the observing conditions. McDonald 2.1\,m observations on $\mathrm{MJD}\,{=}\,59168$ spanning 3\,hr in length do not show this same possible short-term variability to within the $4 \langle A \rangle$ significance threshold (Appendix~\ref{sec:mcd_lcs}). We observe broad Balmer lines in a relatively low-resolution follow-up spectrum from HET (see Figure~\ref{fig:spectra}), and classify ZTF\,J0923+4236 as a DA white dwarf. The lack of emission lines disfavors cataclysmic variable activity as the cause of photometric variability, leaving transiting planetary debris as a likely explanation. Future monitoring is suggested to place stronger constraints on the source of these transit-like features and any metal pollution or circumstellar gas around this white dwarf.

{\bf ZTF\,J034703.18$-$180253.49 (ZTF\,J0347$-$1802)} is the other white dwarf showing likely long-term transits, observed to undergo a $\approx$70-day-long flux dip in its ZTF light curve (Figure~\ref{fig:transits}). Roughly 6.4\,hr of high-speed McDonald photometry both during and outside the flux dip showed no short-term variability, to a limit of 0.87\% (Appendix~\ref{sec:mcd_lcs}). The months-long duration of this transit, while still poorly constrained, likely suggests this object has a very long orbital period. ZTF\,J0347$-$1802 may be unique among these systems by offering a window into the early stages of tidal shredding of a rocky body. Serendipitously, our McDonald follow-up were obtained on $\mathrm{MJD}\,{=}\, 58727$ and $58728$, placing these observations near the deepest portion of the ZTF transit. The calibrated apparent magnitudes from these observations agree with those of ZTF, confirming that the white dwarf got dimmer during this event. We do observe broad hydrogen absorption features in relatively low-resolution follow-up spectroscopy from LDT without any obvious metal pollution (Figure~\ref{fig:spectra}), and classify ZTF\,J0347$-$1802 as a DA white dwarf.

{\bf SDSS\,J010749.38+210745.84 (SDSS\,J0107+2107)} is a metal-polluted DAZ white dwarf with existing SDSS spectra \citep{2015MNRAS.446.4078K} also in our top 1\%. In its ZTF photometry, SDSS\,J0107+2107 continuously exhibits large amplitude scatter (Figure~\ref{fig:example_lcs}). The high-speed photometry we obtained from McDonald corroborates this scatter, revealing essentially continuous transits, many with depths exceeding 25\% in our blue, broad-bandpass {\em BG40} filter. This would make SDSS\,J0107+2107 the coolest white dwarf to exhibit planetary debris transits, with an effective temperature below $8400$\,K (see Table~\ref{tab:transit_params}).

{\bf ZTF\,J032833.52$-$121945.27 (ZTF\,J0328$-$1219)} also reveals essentially continuous transits in its McDonald photometry. Similar to SDSS\,J0107+2107, this object is cool and likely under-luminous in the {\em Gaia} color-magnitude diagram compared to most 0.6\,~$M_{\odot}$ white dwarfs. The short-term variations in the photometry of this object are substantially shallower than those of SDSS\,J0107+2107 and WD\,1145+017. Identification spectroscopy obtained using LDT (see Figure~\ref{fig:spectra}) show deep absorption from the Ca\,{\sc ii}~H \& K lines, and we classify this as a DZ white dwarf.

{\bf WD\,J123432.68+560643.03 (SBSS\,1232+563)} is our least-dramatic candidate for transiting debris, but it is worth brief mention. Although this object did not rank among the top 1\% most variable white dwarfs with the combined {\em Gaia}-ZTF method, it was in the top 1\% using the {\em Gaia}-only metric, as it had \textsc{varindex}~$=~0.025$. We followed this star up with more than 36\,hr of high-speed photometry from McDonald Observatory, one night of which is shown in Figure~\ref{fig:transits}, which revealed shallow, transit-like dips on relatively short timescales. Previous spectroscopy from the SDSS Digital Sky Survey has shown SBSS\,1232+563 to be a metal-polluted white dwarf \citep{2013ApJS..204....5K}, indicating the presence of rocky debris in the circumstellar environment. 

The three systems that appear to undergo transits on short timescales are optimal candidates to follow-up with multi-epoch spectroscopy, especially short-cadence spectrophotometry to probe for substructures within the transiting debris, as previously done for WD\,1145+017 \citep{2018MNRAS.481..703I}. Extended photometric campaigns of these systems are also encouraged to search for repeating patterns and variations in transit shapes, which have been observed to vary subtly on orbit-to-orbit timescales to more dramatically on longer timescales around WD\,1145+017 \citep{Gaensicke2016,Rappaport2016,Rappaport2018}.

\begin{figure}[!t]
    \figurenum{8}
    \epsscale{1.15}
    \plotone{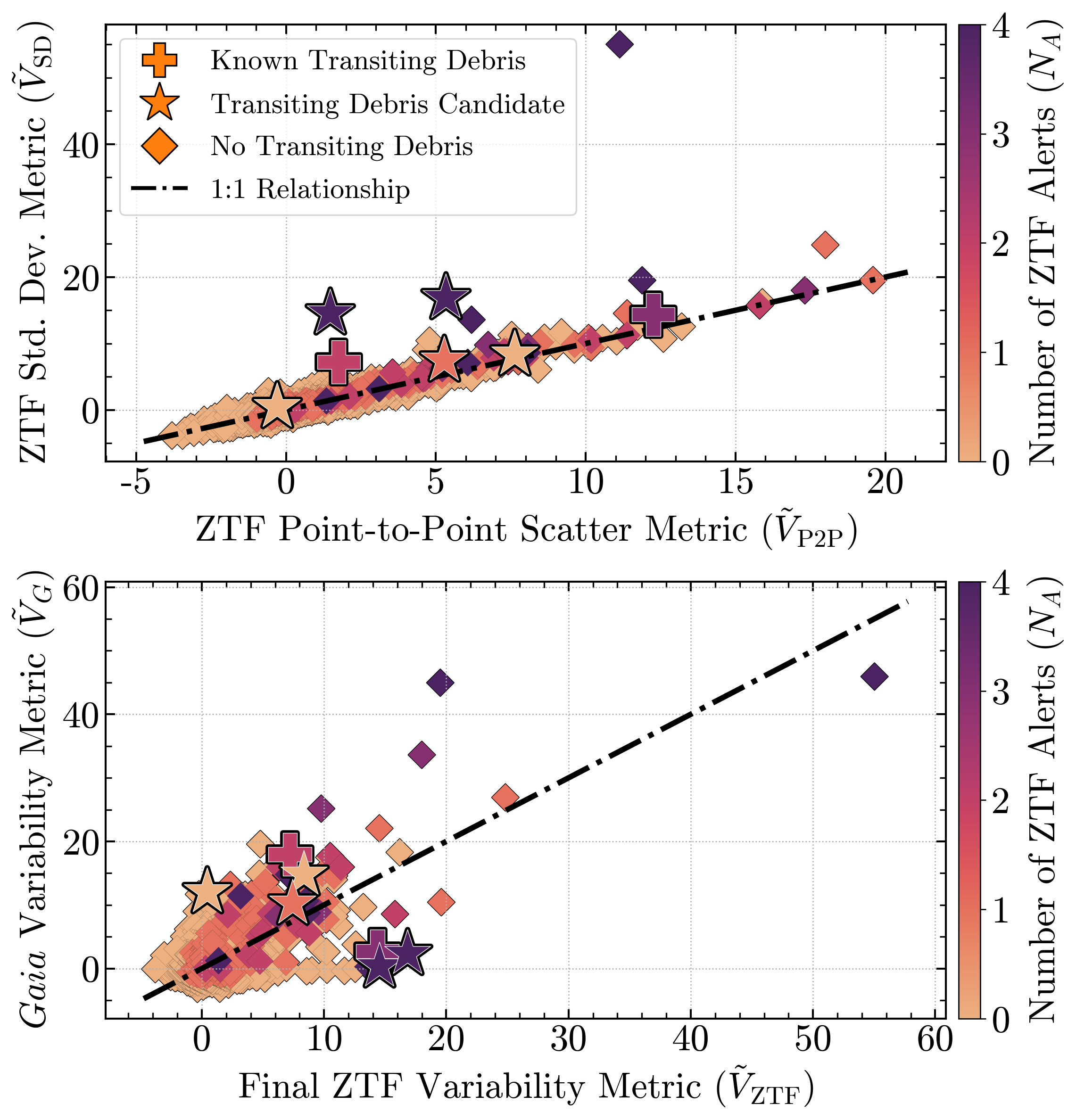}
    \label{fig:sd_vs_p2p}
    \caption{{\bf Top:} Comparison between the detrended ZTF light curve standard deviation metric ($\tilde{V}_{\mathrm{SD}}$) against the point-to-point scatter metric ($\tilde{V}_{\mathrm{P2P}}$). There is generally 1:1 agreement between the two metrics, except for the handful of objects whose longer-term variability dominates, e.g. transits from planetary debris. The objects are color-coded by their ZTF alert metric value, $N_A$. {\bf Bottom:} Comparison between the {\em Gaia} metric ($\tilde{V}_{G}$) vs. the finalized ZTF metric ($\tilde{V}_{\mathrm{ZTF}}$). The regions deviating from the 1:1 agreement are particularly rich areas of interest, evidenced by the detection of two transiting planetary debris system candidates in the $\tilde{V}_{\mathrm{ZTF}} \gg \tilde{V}_G$ regime.}
\end{figure}

Based on the irregular and varied timescales, we show in Figure~\ref{fig:sd_vs_p2p} that the known and candidate transiting debris systems often stand out when comparing different variability metrics, especially comparing the ZTF point-to-point-scatter metric to the ZTF standard-deviation metric. Additionally, the transiting systems occasionally stand out when comparing the final ZTF variability metric to the {\em Gaia} variability metric. This may result from cases where the {\em Gaia} photometry is more aggressively sigma-clipped than the ZTF photometry. Detailed follow-up of objects that stand out in Figure~\ref{fig:sd_vs_p2p} may also make the search for new transiting white dwarfs more efficient. The individual ZTF images of these objects in discrepant regimes, e.g. $\tilde{V}_{\mathrm{ZTF}} \gg \tilde{V}_{G}$, should also be carefully inspected by hand to ensure the discrepancy is not due to artifacts producing false scatter, such as optical ghosts, nearby bright stars, or bad pixel columns.

We conclude by noting one other possible signature to make searches for transiting white dwarfs more efficient: an appearance of being under-luminous in the {\em Gaia} color-magnitude diagram, possibly due to excess circumstellar extinction resulting in erroneous photometric $\log(g)$ inferences. It is noteworthy from Table~\ref{tab:transit_params} that at least three of our new systems have photometrically determined $T_\mathrm{eff}$ and $\log(g)$ and thus overall mass  much larger than the mean mass of field white dwarfs of roughly 0.6\,~$M_{\odot}$ (e.g., \citealt{2016MNRAS.461.2100T}). It may be especially efficient to search for new transiting systems in cases where a star shows both large photometric scatter as well as discrepant spectroscopic and photometric masses, where both are known for a given white dwarf.

Our use of variability metrics from {\em Gaia} DR2 and ZTF has potentially more than tripled the number of white dwarfs that are known to be transited by planetary debris, but we emphasize that we have restricted our search in distance ($<$200\,pc) and to photometric temperatures near the ZZ Ceti instability strip. There is great potential in searching more broadly for white dwarfs with transiting planetary debris using these techniques. We believe that since our standard deviation metric dominates on timescales when the variability periods are much greater than the sampling rate, future targeted applications of this method extended to all white dwarf temperatures and a larger space volume should yield an even larger increase in new candidate transiting debris systems.

%%%%%%%%%%%%%%%%%%%%%%%%%%%%%%%%%%%%%%%%%%%%%%%%%%%%%%%%%%%%%%%%%%%%%%%%%%%%%%%%%%%

                         % New ZZ Cetis              
                
%%%%%%%%%%%%%%%%%%%%%%%%%%%%%%%%%%%%%%%%%%%%%%%%%%%%%%%%%%%%%%%%%%%%%%%%%%%%%%%%%%%

\section{New Confirmed ZZ Cetis}\label{sec:zzcetis}
We report the confirmation of 29 new ZZ Cetis using follow-up high-speed photometry, 9 of which were discovered from our {\em Gaia}-only metric described in Section~\ref{sec:vgaia} and 20 of which were discovered from our {\em Gaia}-ZTF metrics described in Section~\ref{sec:vztf}. This marks our study as the second to use {\em Gaia} DR2 to discover a substantial amount of new ZZ Cetis, following the results of \citet{Vincent2020}. Unlike the \citet{Vincent2020} study, we did not conduct a targeted search of ZZ Cetis, rather we fixated our search on the ZZ Ceti instability strip to use the detection of ZZ Cetis as a proof of concept. We also report the identification of eight more new ZZ Cetis using their ZTF time-series and their inferred placement in the ZZ Ceti instability strip using {\em Gaia} DR2 color information.

\begin{figure}[!t]
\figurenum{9}
\epsscale{1.15}
\plotone{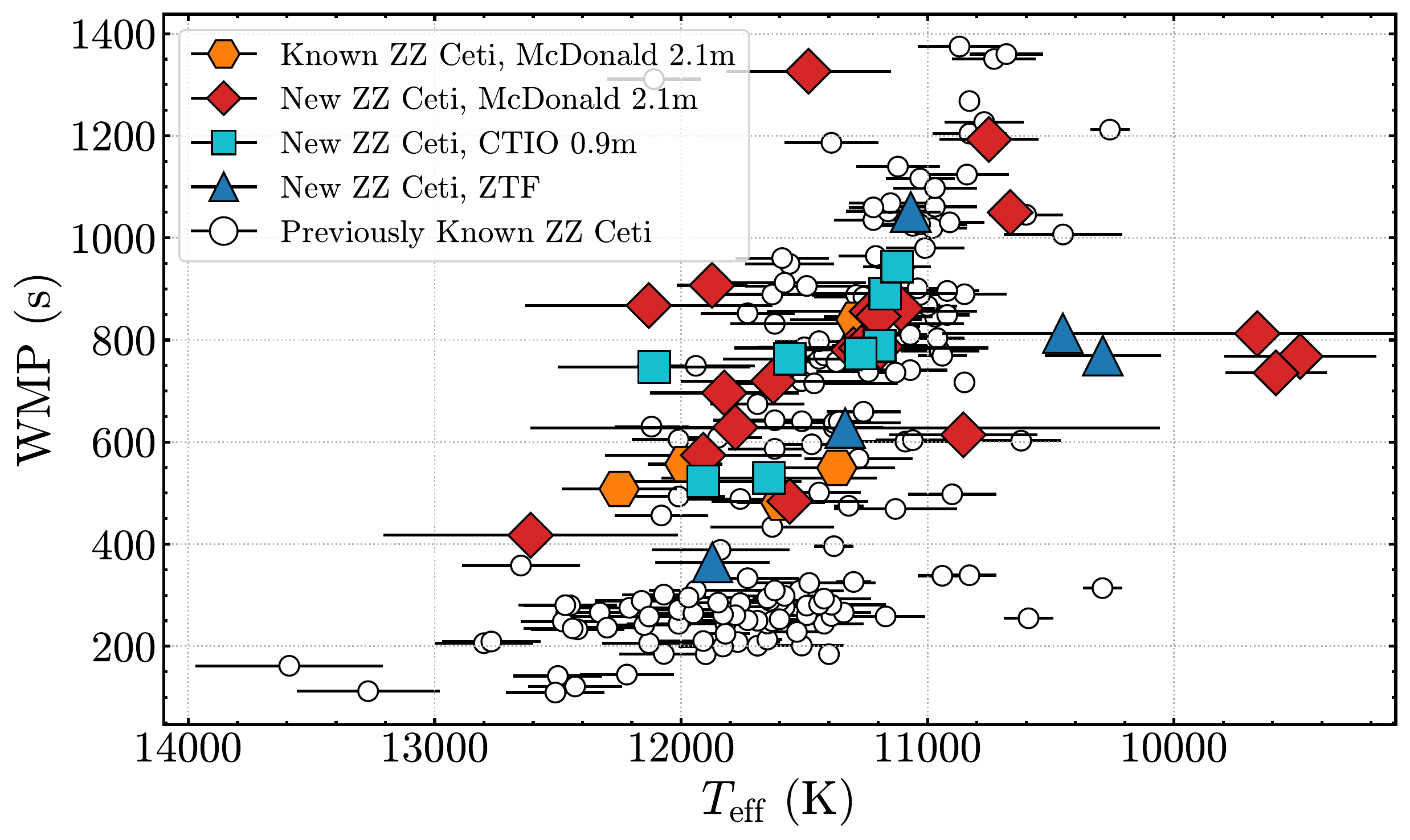}
\label{fig:wmp}
\caption{Distribution of weighted mean pulsation periods (WMP) for independent, significant peaks as a function of photometrically determined effective temperature \citep{GentileFusillo2019}. White circles mark previously analyzed ZZ Cetis \citep{2006ApJ...640..956M,Hermes2017b}, while red diamonds and cyan squares indicate new ZZ Cetis found using this method with follow-up photomtery from the McDonald 2.1\,m and CTIO 0.9\,m, respectively. We show the six known ZZ Cetis from the \citet{Vincent2020} study we observed as orange hexagons and the seven new ZZ Cetis that show significant periods in their ZTF light curves as blue triangles. As pulsating white dwarfs cool their convection zones deepen, driving characteristically longer-period and higher-amplitude pulsations (e.g., \citealt{1993BaltA...2..407C}). Our variability metrics thus preferentially discover cooler pulsating white dwarfs with longer pulsation periods.}
\end{figure}

Since our methods prioritize objects with the most anomalous levels of scatter, most of these new pulsating white dwarfs are characterized by high-amplitude pulsations at long periods. This bias is best demonstrated in Figure~\ref{fig:wmp}, where the weighted mean pulsation periods (WMP) are plotted against $T_{\mathrm{eff}}$. WMP is calculated for a given white dwarf from its linearly independent pulsation period(s), $P$, and corresponding amplitude(s), $A$, using the relationship: $\mathrm{WMP}\,=\,(\sum_i P_i A_i) / (\sum_i A_i)$.

The significant pulsation modes for each ZZ Ceti were determined using the Python package \textsc{Pyriod} \citep{2020AAS...23510606B}\footnote{\url{https://github.com/keatonb/Pyriod}} to conduct a prewhitening routine by hand. We compute the Lomb-Scargle periodogram of each object's McDonald or CTIO time-series and then subtract peaks surpassing the standard 4$\langle A \rangle$ significance threshold until no significant peaks remain in the residual periodogram. We found and removed any combination and harmonic frequencies among the significant peaks to within a frequency tolerance of 0.5 divided by the length of the light curve in seconds in order to determine all the linearly independent, significant pulsation modes present for each object. Only peaks corresponding to periods $80 < P < 2000$\,s were considered, as these bounds form the approximate range for adiabatic non-radial $g$-mode pulsations in ZZ Cetis \citep{Romero2012}. These pulsation spectra are fully detailed in Appendix~\ref{sec:pulsation_spectra}.

Our distribution of WMP in Figure~\ref{fig:wmp} reveals that our method is biased toward finding the highest-amplitude pulsating white dwarfs, which tend to occur in the coolest pulsators with the longest-period pulsation modes. This trend has been noted before in ensemble studies of ZZ Cetis (notably in \citealt{1993BaltA...2..407C} and \citealt{2006ApJ...640..956M}). A more complete search for more pulsating white dwarfs across the entire ZZ Ceti instability strip may require relaxing the threshold that the object must be among the 1\% most variable of all white dwarfs, but would return a more representative distribution like the one attained by \citealt{Vincent2020}.

Notably, we did not discover pulsations in any new extremely low-mass (ELM, $<$0.3\,$M_{\odot}$) white dwarfs in our {\em Gaia}+ZTF sample. This is likely because our temperature selection required targets to have photometrically determined temperatures and surface gravities in \citealt{GentileFusillo2019}, and most ELM white dwarfs do not have values in their DR2 catalog. The space density of ELM white dwarfs is also relatively low, so there are also few within 200\,pc \citep{Kawka2020}. However, we demonstrate that pulsations in ELM white dwarfs can be found from variability metrics using our {\em Gaia}-only sample: WD\,J184424.33+504727.95 ($G=17.9$\,mag) is a new pulsating ELM white dwarf confirmed to exhibit long-period pulsations from McDonald Observatory (Appendix~\ref{sec:mcd_lcs}) consistent with known ELM pulsation periods \citep{2013MNRAS.436.3573H}. We exclude this likely He-core white dwarf from Figure~\ref{fig:wmp}, but its dominant pulsation period is 4220\,s and its WMP exceeds 3900\,s.

Using these variability metrics may also prove fruitful in improving our understanding of the high mass regime of the ZZ Ceti instability strip, as we discovered three new ZZ Cetis with $\log(g)\,{>}\,8.5$. Most notably, WD\,J110604.54+180233.72 (SDSS\,J1106+1802), the most massive of these, has an inferred photometric surface gravity of $\log(g) =$ 8.8 \citep{GentileFusillo2019}, whereas fits to its SDSS spectra imply $\log(g) =$ 9.04 $\pm$ 0.03 \citep{2015MNRAS.446.4078K}. We show in Figure~\ref{fig:example_lcs} and Appendix~\ref{sec:mcd_lcs} that this object undergoes stable, high-amplitude pulsations of order 25\% in peak-to-peak amplitude at a period of 369.7\,s. The amplitude of these pulsations is particularly notable, given higher mass white dwarfs have smaller resonant cavities relative to average mass pulsating white dwarfs, and thus are expected to demonstrate lower amplitude pulsations \citep{Castanheira2010}. Further spectroscopy of this target may also help define the high-mass blue edge of the ZZ Ceti instablity strip: while its photometric temperature is $T_\mathrm{eff} = 12610$\,K \citep{GentileFusillo2019}, its spectroscopic temperature is $T_\mathrm{eff} = 14220 \pm 110$\,K \citep{2015MNRAS.446.4078K}.

Our methods can also reveal other exotic pulsating white dwarfs. Since ZTF is optimized to trigger alerts for transient phenomena, our methods may also be used to detect outbursts from ZZ Cetis in the future. Outbursts were first seen using continuous monitoring from the {\em Kepler} space telescope, and appear to be an internal effect in the coolest pulsating white dwarfs, causing them to undergo a mean flux increase of at least 15\% with durations of order hours \citep{2015ApJ...809...14B, 2015ApJ...810L...5H, 2016ApJ...829...82B}. At least two of our newly confirmed ZZ Cetis, WD\,J203737.79+323833.34 and WD\,J060102.01+541757.82, were observed to increase in relative flux by factors of order $\gtrsim$ 20\% on multiple occasions by ZTF. ZTF also observed WD\,J032438.66+602055.88 to undergo numerous sporadic flux increases exceeding 12\%, which is at least a factor of two increase in amplitude relative to its largest pulsation amplitudes observed from McDonald Observatory. These flux increases could be simply explained by the beating of pulsation modes, though. Still, given the presence of long-period pulsation modes in their follow-up periodograms from McDonald Observatory, these three objects should be closely monitored in the event they undergo future outbursts. 

The transient identification capabilities of ZTF also shed light on WD\,J191852.42+583914.09 (ZTF\,J1918+5839), a newly discovered low-amplitude pulsator confirmed with McDonald photometry, was observed to undergo an enormous flux increase of order $\approx$150\%, slowly decaying over the course of about 10 days in its ZTF DR3 photometry. This would vastly exceed the amplitude and duration of any previously observed outburst in a pulsating white dwarf. J-band photometry from the United Kingdom Infra-Red Telescope (UKIRT) Hemisphere Survey (UHS, \citealt{Dye19}) is consistent with an isolated white dwarf model. Deeper near-infrared imaging could help us rule out a cool line-of-sight companion, since ultracool dwarfs can undergo massive ($\Delta V \sim -10$\,mag) flares \citep[e.g.][]{2019MNRAS.485L.136J}. 

Two additional objects in our top 1\% demonstrated similar phenomena. WD\,J144823.77+572454.62 (ZTF\,J1448+5724) was observed by ZTF to undergo a relative flux increase of $>\,80$\% in $g$ that decayed back to its median flux over the course of 2 days. WD\,J134934.36+280948.93 (ZTF\,J1349+2809) was observed by ZTF to non-monotonically increase from approximately 20\% to over 85\% during six $g$-band observations over the course of about 1.94\,hr. Archived optical photometry and \emph{WISE W1} and \emph{W2} photometry of ZTF\,J13349+2809 are consistent with the \citet{Koester2010Models} DA white dwarf model at the inferred $T_{\mathrm{eff}} = 11{,}250$\,K and $\log(g) = 8.0$ from \citet{GentileFusillo2019}. While ZTF\,J1448+5724 does not possess \emph{WISE} benchmarks, its SED is consistent with the Koester isolated DA white dwarf model at its inferred $T_{\mathrm{eff}} = 10{,}750$\,K and $\log(g) = 7.75$ \citep{GentileFusillo2019}. Notably, both objects have atmospheric parameters (assuming a DA composition) that place them in a  parameter space where ZZ Cetis have been observed to undergo outbursts \citep{Hermes2017b}.

Finally, within the top 1\% sample we identify eight more new ZZ Cetis using the ZTF time-series alone. Seven of these objects were identified after observing significant frequencies in their Lomb-Scargle periodograms beyond their estimated 0.1\% false alarm probability significance thresholds at frequencies greater than $500\,\mu$Hz. For three of these objects, WD\,J062516.34+145558.50, WD\,J143047.25+510730.35, and WD\,J233921.05+512410.79, these detections were aided by multiple nights of dense short-cadence observations that resolved consecutive pulsation peaks. Likewise, despite not showing any significant periodicities, WD\,J182454.39+344331.58 was also densely imaged at short-cadence during two nights in the $r$-band, which resolved variability indicative of ZZ Ceti pulsations. All eight objects have a position in the {\em Gaia} CMD within the ZZ Ceti instability strip, further warranting their classification.

%%%%%%%%%%%%%%%%%%%%%%%%%%%%%%%%%%%%%%%%%%%%%%%%%%%%%%%%%%%%%%%%%%%%%%%%%%%%%%%%%%%

                                %CONCLUSIONS               
                
%%%%%%%%%%%%%%%%%%%%%%%%%%%%%%%%%%%%%%%%%%%%%%%%%%%%%%%%%%%%%%%%%%%%%%%%%%%%%%%%%%%

\section{Conclusions} \label{sec:conclusions}
We have established a novel method capable of identifying variable white dwarfs of numerous types up to $G~\approx~19$\,mag using excess levels of scatter in {\em Gaia} DR2 and ZTF DR3 photometry and ZTF alerts. We applied the metrics to a population of about 12${,}$100 known and candidate white dwarfs within 200\,pc and with temperatures near the ZZ Ceti instability strip, analyzing the top 1\% most variable objects. We fully demonstrate our method with complimentary high-speed photometry from the 2.1\,m Otto Struve telescope at McDonald Observatory, confirming variability in all 33 out of 33 objects we observed in the top 1\%, a 100\% success rate. We also present an easily reproducible global {\em Gaia}-only metric valid over $13 \lesssim G \lesssim 20$ for DR2 photometry. We followed up eight white dwarfs with the Cerro Tololo Inter-American Observatory SMARTS 0.9\,m telescope identified as variable candidates with this technique, confirming ZZ Ceti pulsations in all eight.

Among the previously known and newly confirmed variable white dwarfs we found are mostly pulsating white dwarfs, along with smaller numbers of cataclysmic variables, eclipsing binaries, and transiting debris systems. Our methods recover the only two previously known white dwarfs to show planetary debris transits, and we detect five new candidate transiting debris systems. Given that our target selection was restricted in temperature and distance, we expect to detect many more white dwarfs hosting transiting planetary debris in future applications to forthcoming {\em Gaia} and ZTF data releases, turning these objects into a class we can study to look for differences in debris evolution and dynamics.

While our methods are demonstrated and highly efficient, we acknowledge that they are limited in scope, since we cannot yet confidently classify or identify variability based on excess levels of scatter alone. Follow-up time-series and spectroscopic observations are essential to tapping the full potential of this technique in order to fully characterize the true nature of these objects and systems. We expect that ongoing and upcoming all-sky surveys, and their commitment to public data releases, will provide the community many new exciting variable white dwarfs for years to come.

%%%%%%%%%%%%%%%%%%%%%%%%%%%%%%%%%%%%%%%%%%%%%%%%%%%%%%%%%%%%%%%%%%%%%%%%%%%%%%%%%%%

\acknowledgments
{We thank the anonymous referee for their helpful comments and review of the manuscript. We also recognize Thomas Kupfer for a fruitful discussion on the Zwicky Transient Facility Galactic Plane Survey, which provided the necessary short-cadence observations for some of our detections of new ZZ Cetis using ZTF.

J.A.G. extends gratitude to the Department of Astronomy at the University of Texas at Austin for their awarding of the Summer Undergraduate Research Fellowship that supported this analysis. J.A.G. also acknowledges the Freshman Research Initiative White Dwarf Stream for all the guidance and mentoring it has provided and the doors it has opened.

Z.P.V., D.E.W., and M.H.M. acknowledge support from the United States Department of Energy under grant DE-SC0010623, the National Science Foundation under grant AST-1707419, and the Wootton Center for Astrophysical Plasma Properties under the United States Department of Energy collaborative agreement DE-NA0003843. J.J.H. and I.D.L. acknowledge ground-based observing support through the {\em TESS} Guest Investigator Program Grant 80NSSC19K0378, as well as the {\em K2} Guest Observer program 80NSSC19K0162. B.N.B. and K.A.C. acknowledge funding through the {\em TESS} Guest Investigator Program Grant 80NSSC19K1720 and the National Science Foundation under grant AST-1812874. B.N.B. also acknowledges support from the High Point University Student Government Association, which kindly funded the observing run to the SMARTS 0.9-m at CTIO through bill \#S-18-48. K.J.B.\ is supported by the National Science Foundation under Award AST-1903828. M.H.M. acknowledges support from the NASA ADAP program under grant 80NSSC20K0455. T.M.H. acknowledges support from the National Science Foundation under grant AST-1908119.}

{Follow-up observations obtained using 2.1\,m Otto Struve Telescope and Hobby-Eberly Telescope at The McDonald Observatory of The University of Texas at Austin were vital for this work. This work makes extensive use of publicly archived observations from the Zwicky Transient Facility survey, conducted with the Samuel Oschin 48-inch Telescope at Palomar Observatory. ZTF is a collaboration supported by a consortium of twelve institutions and NSF Grant No. AST-1440341. We also employed data retrieved and processed the European Space Agency (ESA) Gaia mission (https://www.cosmos.esa.int/gaia). All Gaia data processing was orchestrated by the Gaia Data Processing and Analysis Consortium (DPAC, https://www.cosmos.esa.int/web/gaia/dpac/consortium). We queried observations from the Pan-STARRS1 Surveys (PS1) and the PS1 public science archive, which is supported by the consortium of fourteen institutions who make up the PS1 collaboration (https://panstarrs.stsci.edu/) and the NSF Grant No. AST-1238877. These results made use of the Lowell Discovery Telescope at Lowell Observatory. Lowell is a private, non-profit institution dedicated to astrophysical research and public appreciation of astronomy and operates the LDT in partnership with Boston University, the University of Maryland, the University of Toledo, Northern Arizona University and Yale University. Lastly, this research made use of data from the SMARTS 0.9\,m telescope at Cerro Tololo Inter-American Observatory, which is operated as part of the SMARTS Consortium.
}

\facilities{ PO:1.2m (ZTF), Gaia, McD:Struve (ProEM), McD:HET (LRS2), CTIO:0.9m, DCT (DeVeny), ADS, CDS.
}

\software{\textsc{Astropy} \citep{2013A&A...558A..33A, 2018AJ....156..123A},
          \textsc{iraf} (National Optical Astronomy Observatories),
          \textsc{lmfit} \citep{newville_matthew_2014_11813},
          \textsc{matplotlib} \citep{Matplotlib2007},
          \textsc{Numpy} \citep{numpy2020}
          \textsc{pandas} \citep{pandas2020},
          \textsc{phot2lc} (\url{https://github.com/zvanderbosch/phot2lc}),
          \textsc{Photutils} \citep{Bradley2020},
          \textsc{Pyriod} (\url{https://github.com/keatonb/Pyriod}),
          CDS's (Strasbourg, France) SIMBAD and VizieR online pages and tables,
          and the NASA Astrophysics Data System (ADS) repositories.
          }

%%%%%%%%%%%%%%%%%%%%%%%%%%%%%%%%%%%%%%%%%%%%%%%%%%%%%%%%%%%%%%%%%%%%%%%%%%%%%%%%%

\bibliography{references}{}
\bibliographystyle{aasjournal}

%%%%%%%%%%%%%%%%%%%%%%%%%%%%%%%%%%%%%%%%%%%%%%%%%%%%%%%%%%%%%%%%%%%%%%%%%%%%%%%%

\appendix

\section{{\em Gaia} Astrometric Quality Cuts}\label{sec:gaiacuts}

Below we show all the Astronomical Data Query Language (ADQL) astrometric quality cuts we imposed on the \citet{GentileFusillo2019} {\em Gaia} DR2 catalog, as recommended in \citet{2018A&A...616A...2L} and \citet{2018A&A...616A...4E}:

\textsc{visibility\_periods\_used} $> 8$ \& \textsc{parallax} $> 5$ \& \textsc{parallax\_over\_error} $> 10$ \& \textsc{astrometric\_excess\_noise} $< 1.0$ \& \textsc{phot\_proc\_mode} $<$ 0.1 \& \textsc{phot\_bp\_mean\_flux\_over\_error} $> 10$ \& \textsc{phot\_rp\_mean\_flux\_over\_error $>$ 10} \& \textsc{phot\_g\_mean\_flux\_over\_error} $> 20$ \& \textsc{phot\_bp\_rp\_excess\_factor} $< 1.3+0.06$*pow(\textsc{phot\_bp\_mean\_mag} $-$ \textsc{phot\_rp\_mean\_mag},2) \& \textsc{phot\_bp\_rp\_excess\_factor} $> 1.0+0.015$*pow(\textsc{phot\_bp\_mean\_mag} $-$ \textsc{phot\_rp\_mean\_mag},2) \& \textsc{astrometric\_chi2\_al} / (\textsc{astrometric\_n\_good\_obs\_al} $-5$) $<$ 1.44*maxReal(1,exp($-0.4$*(\textsc{phot\_g\_mean\_mag} $-$ 19.5))) \& (\textsc{phot\_g\_mean\_mag} $+$ 5 * log10(\textsc{parallax}/100)) $>$ ((4.0 * \textsc{bp\_rp}) $+$ 9.0)

%%%%%%%%%%%%%%%%%%%%%%%%%%%%%%%%%%%%%%%%%%%%%%%%%%%%%%%%%%%%%%%%%%%%%%%%%%%%%%%

\section{Pan-STARRS Decontamination Criteria}\label{sec:ps1criteria}
Below we present the decontamination criteria we imposed on our astrometrically clean sample of 18${,}$269 objects using Pan-STARRS PS1 catalog. First, we require that objects must have real (i.e., not \textsc{NaN}) \textsc{gMeanPSFMag} and \textsc{rMeanPSFMag} values. If one of these are \textsc{NaN}s, and the next nearest measurements are within 1.75$^{\prime \prime}$ with $|\textsc{gMeanPSFMag} - G_{BP}| \leq 0.15$\, mag and $|\textsc{rMeanPSFMag} - G| \leq 0.15$\,mag, then those values are used for the object. Otherwise the object is flagged and removed from the analysis.

Assuming an average FWHM seeing of 2.5$^{\prime \prime}$ in the ZTF photometry, the remaining decontamination criteria are:
\begin{itemize}
    \setlength\itemsep{0pt}
    \setlength\parskip{5pt}
    \item Objects with a neighboring star within 5.0$^{\prime \prime}$ are flagged unless the source is at least 2.0 magnitudes dimmer
    \item Objects with a neighboring star between 5.0 - 7.5$^{\prime \prime}$ are flagged unless the source is at least 1.0 magnitudes dimmer
    \item Objects with a neighboring star more than 2.0 magnitudes brighter between 7.5 - 12.0$^{\prime \prime}$ are flagged
    \item Objects with a 13th magnitude star or brighter within 30.0$^{\prime \prime}$ are flagged
    \item Objects with a 10th magnitude star or brighter within 60.0$^{\prime \prime}$ are flagged
\end{itemize}

%%%%%%%%%%%%%%%%%%%%%%%%%%%%%%%%%%%%%%%%%%%%%%%%%%%%%%%%%%%%%%%%%%%%%%%%%%%%%%%

\section{McDonald 2.1\,m Light Curves and Periodograms}\label{sec:mcd_lcs}

\begin{figure}[H]
    \centering
    \figurenum{10}
    \epsscale{1.1}
    \plotone{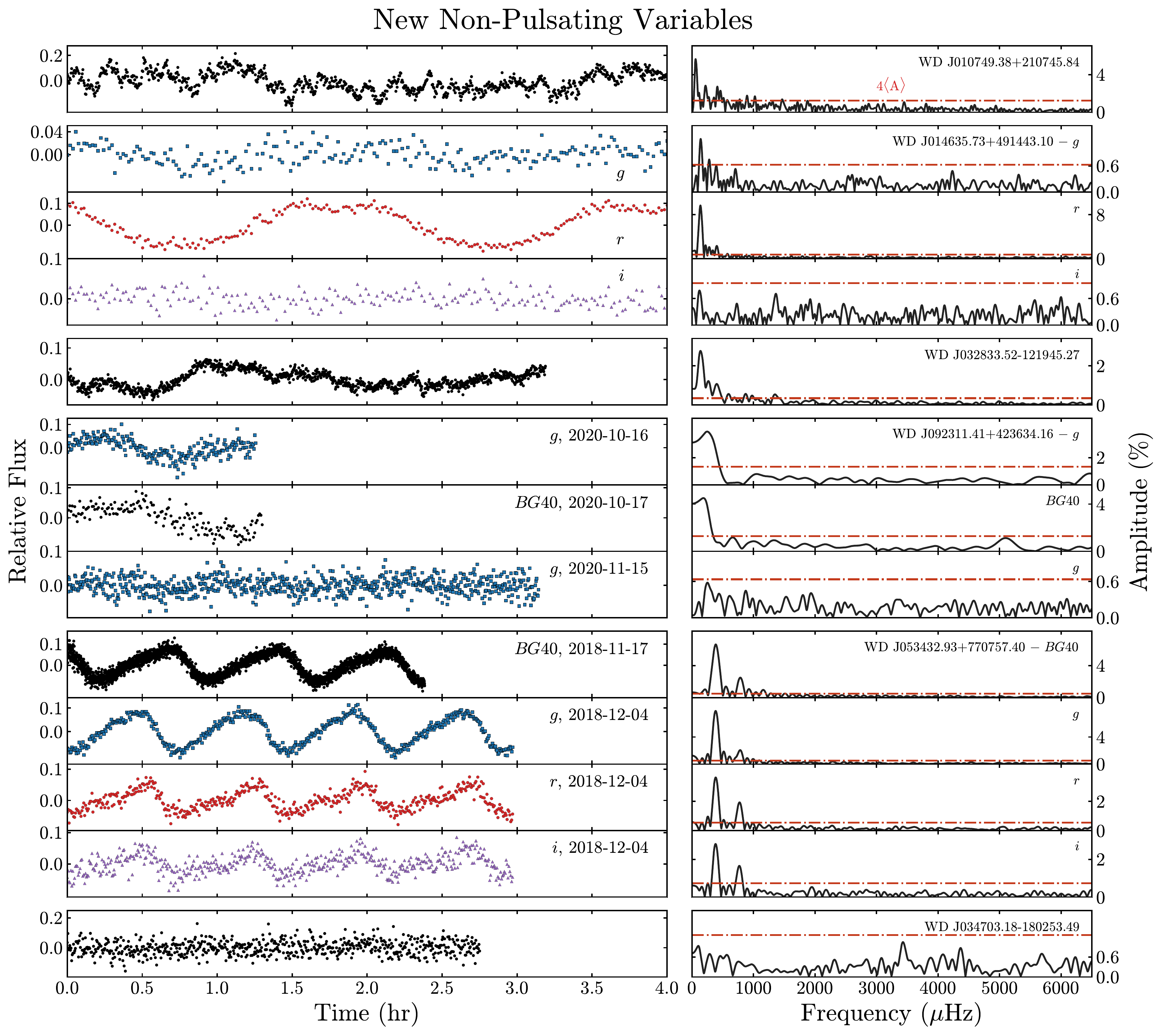}
    \label{fig:mcd_non_puls}
    \caption{McDonald 2.1\,m high-speed time-series photometry for the six new non-pulsating variable objects we followed up in the top 1\% using {\em Gaia}+ZTF. Black dots represent observations taken in the {\em BG40} filter, blue squares represent SDSS $g$, red circles represent SDSS $r$, and purple triangles represent SDSS $i$. Light curves from different nights are noted. Although these photometry show ZTF\,J0347-1802 not to be significantly variable on short timescales, we clearly see a transit in its ZTF photometry that implies the presence of long-term variability. ZTF\,J0923+4236 does appear to exhibit short-term variability. Due to short observing windows at the time of observing, we are uncertain of the origin of this variability. Similarly, the origin of the consistent, stable pattern shown in the multi-color photometry of ZTF\,J0534+7707 is unknown, with the most likely explanation being a surface inhomogeneity like a dark spot due to a strong magnetic field.}
\end{figure}

\begin{figure}[H]
    \centering
    \epsscale{1.1}
    \figurenum{11}
    \plotone{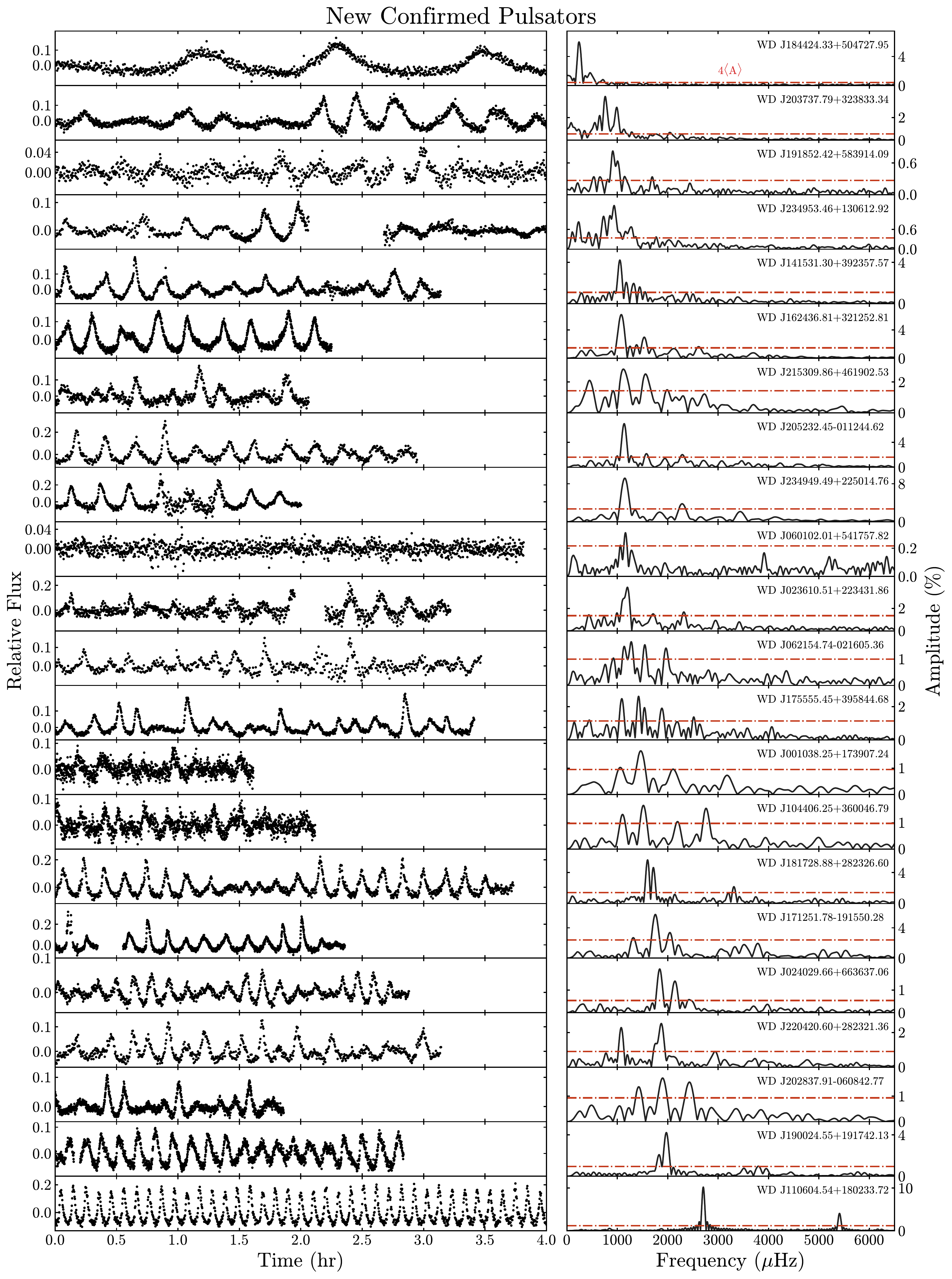}
    \label{fig:mcd_davs}
    \caption{McDonald 2.1\,m high-speed time-series photometry for the 21 new confirmed ZZ Cetis in the top 1\% using {\em Gaia}+ZTF. We also show the photometry for WD\,J1844+5047, the new ELMV we report using the {\em Gaia}-only technique.}
\end{figure}

\begin{figure}[H]
    \centering
    \epsscale{1.1}
    \figurenum{12}
    \plotone{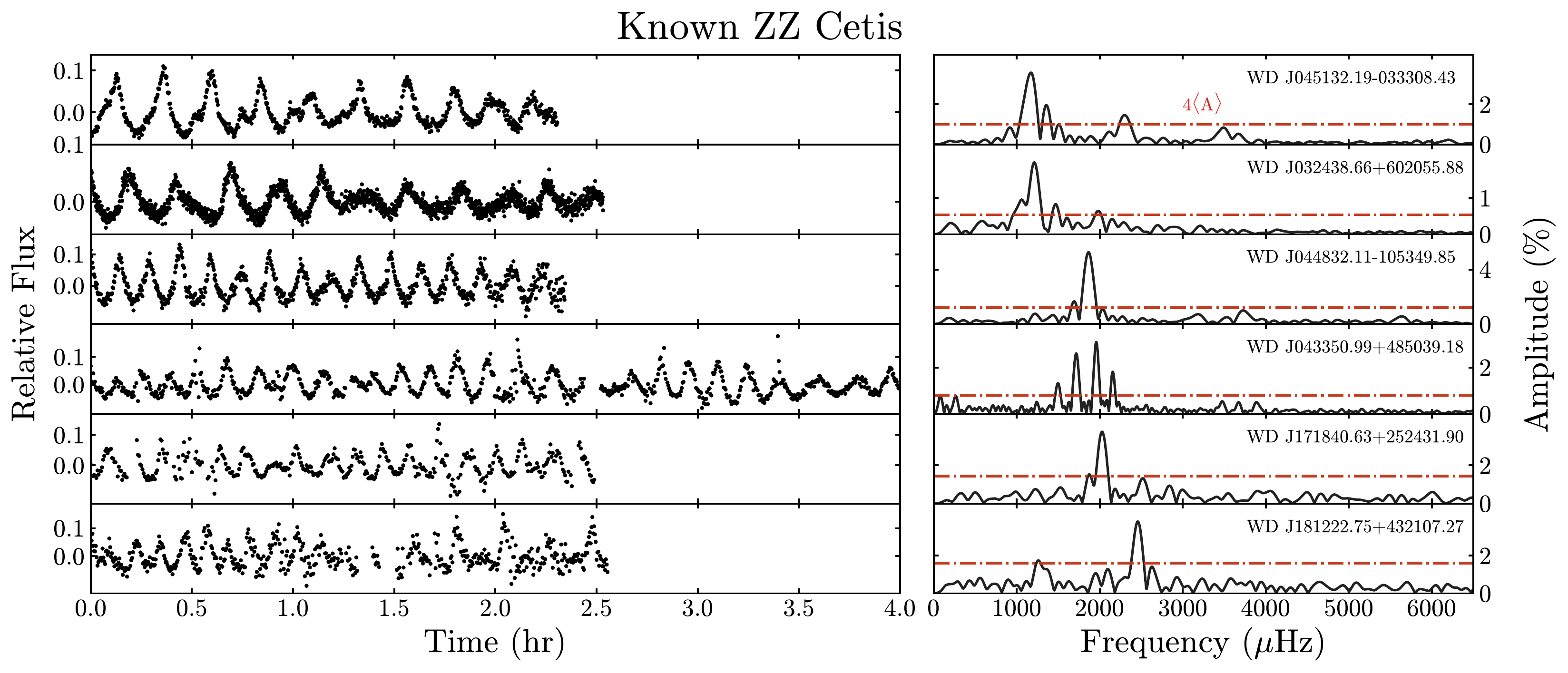}
    \caption{McDonald 2.1\,m high-speed time-series photometry for six recently discovered ZZ Cetis in \citet{Vincent2020} residing in the top 1\% that we independently verified.}
    \label{fig:mcd_known_davs}
\end{figure}

In this Appendix we show the light curves and Lomb-Scargle periodograms from our follow-up campaign using the McDonald Observatory 2.1\,m Otto Struve telescope. Figure~\ref{fig:mcd_non_puls} displays all six confirmed non-pulsating variables. Figure~\ref{fig:mcd_known_davs} shows our observations of the six newly announced ZZ Cetis in \citet{Vincent2020} that we followed-up independently. Finally, Figure~\ref{fig:mcd_davs} shows the new ZZ Cetis we announce in this work, including the extremely low mass pulsator WD\,J1844+5047 found using the {\em Gaia}-only method. All objects were observed through a {\em BG40} blue-bandpass filter unless specified otherwise.

\newpage

%%%%%%%%%%%%%%%%%%%%%%%%%%%%%%%%%%%%%%%%%%%%%%%%%%%%%%%%%%%%%%%%%%%%%%%%%%%%%%%

\section{CTIO SMARTS 0.9\,m Light Curves and Periodograms}\label{sec:ctio_lcs}

\begin{figure}[H]
    \centering
    \epsscale{1.1}
    \figurenum{13}
    \plotone{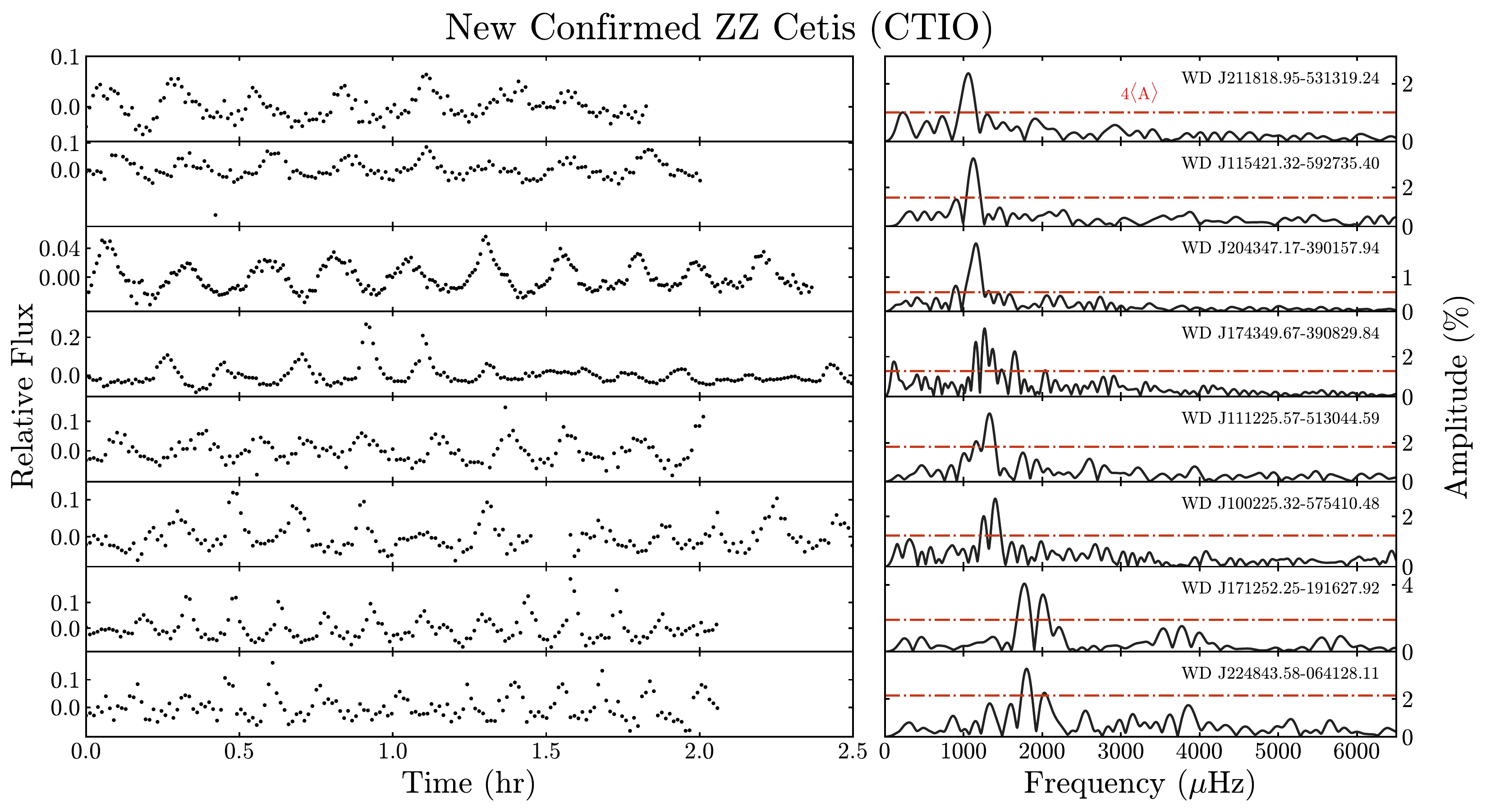}
    \caption{SMARTS CTIO 0.9\,m high-speed, time-series photometry for the eight new ZZ Cetis found using the {\em Gaia}-only method.}
    \label{fig:ctio_zzcetis}
\end{figure}

In this Appendix we show the light curves and Lomb-Scargle periodograms from our follow-up campaign using the SMARTS 0.9\,m telescope at Cerro Tololo Inter-American Observatory. Figure~\ref{fig:ctio_zzcetis} shows the eight new ZZ Cetis we found using the {\em Gaia}-only method. The stellar parameters of these objects are shown in Table~\ref{tab:ctio}. Note that photometry of WD\,J1712$-$1915, an object we observed at McDonald Observatory, is also shown here.

\begin{deluxetable}{cccccccccccc}[H]
\tablecaption{Table of parameters and metrics for new ZZ Cetis selected from the top 1\% most variable white dwarfs from our {\em Gaia}-only variability search. The parameters here are sourced from the \citet{GentileFusillo2019} catalog with $\alpha$ and $\delta$ representing the J2015.5 epoch.}
\tablewidth{0pt}
\tablenum{4}
\label{tab:ctio}
\tablehead{
\colhead{WD} & \colhead{$\alpha$ (deg)} & \colhead{$\delta$ (deg)} & \colhead{$G$} & \colhead{$T_{\mathrm{eff}}$} & \colhead{$\log(g)$} & \colhead{\textsc{varindex}}
}
\startdata
 WD\,J100223.96$-$575507.91  & 150.59947 & $-$57.91868 &  16.8 &   12110 &  8.0 & 0.0130 \\
 WD\,J111221.43$-$513003.90  & 168.08948 & $-$51.50112 &  16.4 &   11190 &  8.0 & 0.0121 \\
 WD\,J115414.55$-$592658.81  & 178.56046 & $-$59.44960 &  16.9 &   11170 &  8.0 & 0.0100 \\
 WD\,J171251.78$-$191550.28  & 258.21559 & $-$19.26396 &  16.3 &   11910 &  8.1 & 0.0261 \\
 WD\,J174349.28$-$390825.95  & 265.95468 & $-$39.14034 &  13.6 &   11560 &  8.1 & 0.0246 \\
 WD\,J204349.21$-$390318.02  & 310.95500 & $-$39.05652 &  13.8 &   11270 &  8.0 & 0.0063 \\
 WD\,J211815.52$-$531322.72  & 319.56444 & $-$53.22334 &  15.9 &   11120 &  7.9 & 0.0086 \\
 WD\,J224840.07$-$064244.65  & 342.16694 &  $-$6.71254 &  16.9 &   11640 &  8.1 & 0.0257 \\
\enddata
\end{deluxetable}

%%%%%%%%%%%%%%%%%%%%%%%%%%%%%%%%%%%%%%%%%%%%%%%%%%%%%%%%%%%%%%%%%%%%%%%%%%%%%%%

\section{The {\em Gaia} Variability Metric as a Global Proxy for Variability}\label{sec:varindex}

Our listed catalog of variable sources in Appendix~\ref{sec:top1table} covers only the sky with overlapping {\em Gaia} and ZTF coverage, excluding most of the southern hemisphere. We discuss here an extension of our construction of an all-sky variability index from \citet{Hermes2018} using only data from {\em Gaia} DR2, expanding upon the discussion in Section~\ref{sec:vgaia}.

First, we fit an exponential function to our {\em Gaia} variability metric, described in Section~\ref{sec:vgaia}, which quantifies {\em Gaia} photometric scatter: $V_G = \sigma_G \sqrt{n_{\mathrm{obs},G}} / \langle G \rangle $, where $\sigma_G$ is the value \textsc{phot\_g\_mean\_flux\_error}, $\langle G \rangle$ is the value \textsc{phot\_g\_mean\_flux}, and $n_{\mathrm{obs},G}$ is the value \textsc{phot\_g\_n\_obs} in the {\em Gaia} DR2 catalog.

\begin{figure*}[!b]
    \centering
    \epsscale{1.1}
    \figurenum{14}
    \plotone{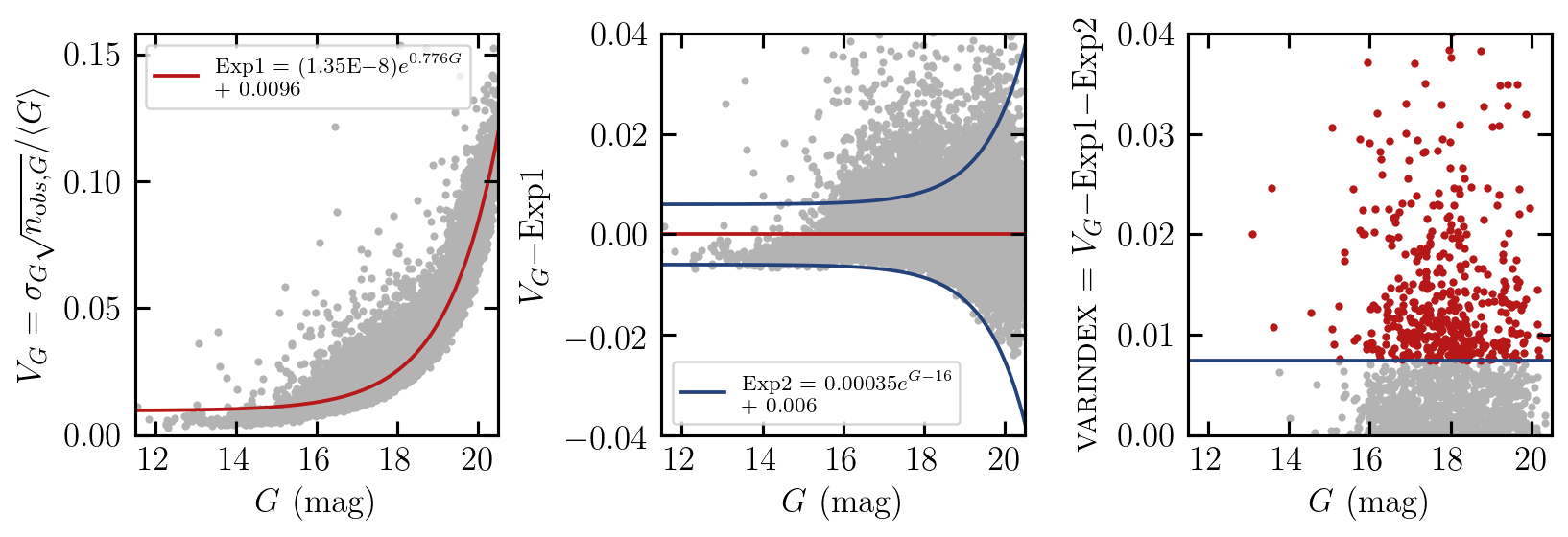}
    \caption{{\bf Left}: Empirical {\em Gaia} DR2 variability metric as a function of $G$-band magnitude, and define our first exponential fit in red. {\bf Middle}: Residuals of the first exponential fit, which does not entirely capture the photometric scatter at the faintest magnitudes. A second exponential fit, which kicks in at $G=16$\,mag, is shown in blue, fitted by eye to the negative residuals. {\bf Right}: The final \textsc{varindex} values from the residuals of the two exponential functions, defined in Equation~\ref{eqn:varindex}. The 1\% most variable objects have \textsc{varindex} $>0.0074$ and are the objects also marked in red in Figure~\ref{fig:gaia_only_variables}.}
    \label{fig:GaiaVarAppendix}
\end{figure*}

The first exponential fit is shown with the red line in the left-most panel in Figure~\ref{fig:GaiaVarAppendix}: Exp1 = ($1.35\times10^{-8}$)$e^{0.776G} + 0.0096$. In most cases, objects with {\em Gaia} variability metrics far from this single exponential fit are excellent candidate variables.

However, we noticed that the residuals to this single exponential function show excess scatter at the faintest magnitudes, shown in the middle panel of Figure~\ref{fig:GaiaVarAppendix}. This could incorrectly flag some of the faintest white dwarfs as variable. Therefore we used an additional exponential function that kicked in at magnitudes fainter than $G>16$\,mag in order to follow the bounds of the negative residuals. This choice at $G=16.0$\,mag was motivated by the convergence of the calibrations of the large–scale and a small–scale components of the photometric calibration in {\em Gaia} DR2 (see especially Figure 9 of \citealt{2018A&A...616A...3R}). This second exponential function was fit by-eye to the envelope of the unphysical negative residuals, which are only tracing the shape of empirical photometric variability at the faintest magnitudes. This second exponential function is mirrored and shown in blue in the middle panel of Figure~\ref{fig:GaiaVarAppendix}: Exp2 = 0.00035$e^{G-16.0} + 0.006$.

Subtracting both exponential functions from the {\em Gaia} variability metric yields for us a final \textsc{varindex} value, the largest of which ranks for us the highest-probability variables at all magnitudes for white dwarfs within 200\,pc in {\em Gaia} DR2. The top 1\% most variable white dwarfs are shown in the right-most panel of Figure~\ref{fig:GaiaVarAppendix}; the 1\% most variable objects have \textsc{varindex} $>0.0074$. Our double-exponential calibration should define any object with \textsc{varindex} $>0.0$ as a strong candidate for variability (see discussion in Section~\ref{sec:gaia_results}).

We do not include a catalog of these most-variable white dwarfs from only {\em Gaia} DR2 photometry, but instead simply include here the relationship for future researchers to directly duplicate so that they can curate their own variability catalogs from the {\em Gaia} data directly:

\begin{equation}
\begin{split}
    \textsc{varindex$_{\mathrm{DR2}}$} = (\textsc{phot\_g\_mean\_flux\_error}  \sqrt{\textsc{phot\_g\_n\_obs}} / \textsc{phot\_g\_mean\_flux}) \\ - (1.35\times10^{-8}e^{0.776\textsc{phot\_g\_mean\_mag}} + 0.00035e^{\textsc{phot\_g\_mean\_mag}-16.0} + 0.0156)
    \label{eqn:varindex}
\end{split}
\end{equation}

We conclude with the comment that this relationship and variability catalogs are calibrated for {\em Gaia} DR2, and will change slightly with future data releases. We have investigated how this relationship is modified by new photometry and astrometry from the {\em Gaia} early data release (eDR3) made public on 2020 December 3 \citep{2020arXiv201201533G}.

First, we confirm that the majority of white dwarfs from DR2 that were among the top 1\% most variable are also in the top 1\% using the eDR3 values for their photometric uncertainties. We have further recalibrated our exponential fits with the new eDR3 photometry, using an identical method to the DR2 procedure outlined in this Appendix. That yields the following updated function:

\begin{equation}
\begin{split}
    \textsc{varindex$_{\mathrm{eDR3}}$} = (\textsc{phot\_g\_mean\_flux\_error}  \sqrt{\textsc{phot\_g\_n\_obs}} / \textsc{phot\_g\_mean\_flux}) \\ - (8.31\times10^{-9}e^{0.794\textsc{phot\_g\_mean\_mag}} + 0.0005e^{\textsc{phot\_g\_mean\_mag}-17.0} + 0.019)
    \label{eqn:varindex_edr3}
\end{split}
\end{equation}

Of the $44{,}045$ white dwarfs that are within 200\,pc both according to {\em Gaia} DR2 and eDR3, the top 1\% (top 440) have \textsc{varindex$_{\mathrm{eDR3}}$} $>0.0027$. Of those 440 WDs, 284 (65\%) were also in the top 1\% most variable with the DR2 photometry, and 27 are previously known pulsators. We defer to future work further exploration of the {\em Gaia} eDR3 \textsc{varindex}.

%%%%%%%%%%%%%%%%%%%%%%%%%%%%%%%%%%%%%%%%%%%%%%%%%%%%%%%%%%%%%%%%%%%%%%%%%%%%%%%

\section{Table of Top 1\% Variable Objects}\label{sec:top1table}

In Table~\ref{tab:top1} we show the stellar parameters and and metric values for all the 121 objects in our combined {\em Gaia} and ZTF top 1\%. A description of columns is found in the note below.

\startlongtable
\begin{deluxetable}{cccccccccccc}
\tablenum{5}
\tablecaption{Table of parameters and metrics for top 1\%. }
\label{tab:top1}
\tablewidth{0pt}
\tablecomments{Description of columns: all of the object names and stellar parameters are directly sourced from the \citet{GentileFusillo2019} {\em Gaia} DR2 catalog for candidate white dwarfs. Here, $\alpha$ and $\delta$ are the J2015.5 {\em Gaia} right ascension and declination in degrees, $G$ is the average observed {\em Gaia} G-band magnitude, and $T_{\mathrm{eff}} \ (\mathrm{K})$  and $\log(g) \ (\mathrm{dex})$ are derived from the {\em Gaia} photometry assuming a pure hydrogen photosphere. The rank parameter, $R$, and individual variability metric values for each object are also presented. We use the Observatory column to indicate which objects were followed up with the McDonald Observatory 2.1\,m telescope (McD, see Appendix~\ref{sec:mcd_lcs}), the CTIO SMARTS 0.9\,m (CTIO, see Appendix~\ref{sec:ctio_lcs}), and the eight new ZZ Cetis classified using their ZTF light curves (ZTF). The key for these classifications is as follows: ZZ = known ZZ Ceti, cZZ = new confirmed ZZ Ceti, CV = known cataclysmic variable, cCV = new confirmed cataclysmic variable, TR = known transiting debris, cTR = new candidate white dwarf with transiting debris, EB = known eclipsing binary, MS = rotational modulation due to a cool magnetic spot, and V are objects whose variability has yet to be classified or confirmed.}
\tablehead{
\colhead{WD} & \colhead{$\alpha$ (deg)} & \colhead{$\delta$ (deg)} & \colhead{Class} & \colhead{$R$} & \colhead{$\tilde{V}_{\mathrm{ZTF}}$} & \colhead{$\tilde{V}_{G}$} & \colhead{$N_{A}$} & \colhead{$G$} & \colhead{$T_{\mathrm{eff}}$} & \colhead{$\log(g)$} & \colhead{Observed}
}
\startdata
 WD\,J001038.25+173907.24    &  2.65952 &  17.65175 & cZZ   &  27.7 &  4.1 &  9.7 &  1 & 17.8 &   11220 &  7.9 & McD                \\
 WD\,J002511.11+121712.39    &  6.29599 &  12.28661 & CV    & 139.7 &  9.8 & 25.2 &  3 & 17.5 &    8910 &  7.3 & \\
 WD\,J002535.80+223741.89    &  6.39950 &  22.62813 & V     &  15.2 &  4.1 &  3.5 &  1 & 18.0 &   11490 &  8.1 & \\
 WD\,J004711.37+305609.18    & 11.79746 &  30.93552 & cZZ     &  20.1 &  3.9 &  6.1 &  1 & 17.6 &   10450 &  7.5 & ZTF \\
 WD\,J010207.20$-$003259.57  & 15.53151 &  -0.55041 & ZZ    &  14.6 &  3.3 & 11.3 &  0 & 18.2 &   10320 &  7.9 & \\
 WD\,J010528.63+020501.63    & 16.37012 &   2.08360 & ZZ    &  15.3 &  5.6 &  9.7 &  0 & 16.7 &   11180 &  7.9 & \\
 WD\,J010749.38+210745.84    & 16.95550 &  21.12910 & cTR   &  35.3 &  7.4 & 10.2 &  1 & 19.2 &    7590 &  8.8 & McD                \\
 WD\,J013906.17+524536.89    & 24.77633 &  52.76027 & TR    &  75.5 &  7.2 & 17.9 &  2 & 18.5 &    9420 &  7.9 & McD                \\
 WD\,J014635.73+491443.10    & 26.64891 &  49.24527 & cCV   &  51.3 &  8.0 &  9.1 &  2 & 17.0 &    8550 &  8.3 & McD \\
 WD\,J014721.82$-$215651.39  & 26.84125 & -21.94771 & ZZ    &  73.1 & 15.8 &  8.6 &  2 & 15.2 &   10840 &  8.0 & \\
 WD\,J020158.87$-$105438.25  & 30.49558 & -10.91064 & ZZ    &  24.3 &  4.8 & 19.6 &  0 & 16.9 &   11250 &  8.1 & \\
 WD\,J021155.07+190631.87    & 32.97941 &  19.10887 & V     &  14.9 &  5.1 &  9.8 &  0 & 18.2 &   11430 &  8.1 & \\
 WD\,J022823.31$-$134726.71  & 37.09745 & -13.79091 & V     &  50.8 & 10.8 & 14.6 &  1 & 17.0 &   11530 &  8.0 & \\
 WD\,J023214.00+344304.51    & 38.05832 &  34.71759 & V     &  32.2 &  5.0 &  5.8 &  2 & 17.0 &   11230 &  8.0 & \\
 WD\,J023610.51+223431.86    & 39.04377 &  22.57553 & cZZ   &  23.0 &  2.4 &  9.2 &  1 & 18.4 &    9660 &  7.7 & McD                \\
 WD\,J024029.66+663637.06    & 40.12259 &  66.61015 & cZZ   &  32.1 &  6.5 &  9.5 &  1 & 15.6 &   11870 &  8.0 & McD                \\
 WD\,J024927.57+325112.43    & 42.36522 &  32.85323 & ZZ    &  17.3 &  6.2 & 11.1 &  0 & 16.1 &   11390 &  8.0 & \\
 WD\,J030648.35$-$172332.93  & 46.70206 & -17.39228 & V     &  22.1 &  8.5 & 13.7 &  0 & 16.7 &   10990 &  8.0 & \\
 WD\,J032438.66+602055.88    & 51.16129 &  60.34886 & ZZ    &  17.9 &  3.8 &  2.1 &  2 & 16.1 &   11250 &  8.0 & McD                \\
 WD\,J032833.52$-$121945.27  & 52.14013 & -12.32930 & cTR   &  23.1 &  8.4 & 14.8 &  0 & 16.6 &    8750 &  8.5 & McD                \\
 WD\,J034703.18$-$180253.49  & 56.76399 & -18.04825 & cTR   &  74.4 & 14.5 &  0.3 &  4 & 17.4 &   13370 &  8.9 & McD                \\
 WD\,J034706.79$-$115808.89  & 56.77869 & -11.96926 & ZZ    &  20.2 & 11.0 &  9.1 &  0 & 16.0 &   11670 &  8.1 & \\
 WD\,J035454.20+074608.59    & 58.72630 &   7.76831 & V     &  16.5 &  7.0 &  9.5 &  0 & 16.6 &   11190 &  8.0 & \\
 WD\,J041856.64+271748.31    & 64.73629 &  27.29644 & ZZ    &  60.0 & 19.6 & 10.4 &  1 & 15.1 &   12380 &  7.8 & \\
 WD\,J042017.25+361627.27    & 65.07199 &  36.27301 & ZZ    &  22.9 & 13.2 &  9.7 &  0 & 15.7 &   11270 &  7.9 & \\
 WD\,J043139.22+192127.77    & 67.91356 &  19.35762 & V     &  16.3 &  3.3 &  4.8 &  1 & 17.5 &   12560 &  8.1 & \\
 WD\,J043350.99+485039.18    & 68.46260 &  48.84424 & ZZ    &  49.7 &  9.9 & 15.0 &  1 & 15.9 &   11370 &  8.0 & McD                \\
 WD\,J044258.31+323715.63    & 70.74314 &  32.62089 & V     &  41.2 & 10.2 & 10.5 &  1 & 17.4 &   11640 &  7.9 & \\
 WD\,J044832.11$-$105349.85  & 72.13361 & -10.89725 & ZZ    &  57.7 &  6.2 &  8.2 &  3 & 16.3 &   12250 &  8.5 & McD                \\
 WD\,J045132.19$-$033308.43  & 72.88438 &  -3.55222 & ZZ    &  82.0 & 11.4 & 16.0 &  2 & 16.1 &   11290 &  8.0 & McD                \\
 WD\,J045927.24+552521.05    & 74.86189 &  55.42159 & ZZ    &  24.4 &  4.8 & 19.6 &  0 & 16.0 &   11840 &  8.0 & \\
 WD\,J050639.84$-$140511.02  & 76.66592 & -14.08638 & V     &  13.0 &  9.6 &  3.4 &  0 & 18.2 &   11180 &  7.9 & \\
 WD\,J050932.26+450954.98    & 77.38430 &  45.16515 & V     &  24.1 &  5.4 &  6.7 &  1 & 17.5 &   11030 &  7.9 & \\
 WD\,J051013.52+043855.13    & 77.55636 &   4.64821 & ZZ    &  14.7 &  8.0 &  6.7 &  0 & 15.4 &   11780 &  8.1 & \\
 WD\,J053349.69+155708.02    & 83.45700 &  15.95194 & V     &  22.2 &  4.5 &  6.6 &  1 & 17.6 &   12020 &  8.0 & \\
 WD\,J053432.93+770757.40    & 83.63753 &  77.13198 & MS     &  22.6 &  9.1 & 13.5 &  0 & 16.5 &   10410 &  8.3 & McD \\
 WD\,J060102.01+541757.82    & 90.25830 &  54.29892 & cZZ   &  17.7 &  4.8 &  1.2 &  2 & 16.7 &   11110 &  8.1 & McD                \\
 WD\,J062154.74$-$021605.36  & 95.47794 &  -2.26792 & cZZ   &  23.9 &  4.1 &  7.9 &  1 & 17.4 &    9490 &  7.6 & McD            \\
 WD\,J062516.34+145558.50    & 96.31799 &  14.93281 & cZZ     &  42.9 &  5.6 &  8.7 &  2 & 17.5 &   11280 &  8.1 & ZTF \\
 WD\,J062555.04$-$141442.31  & 96.47949 & -14.24532 & V     &  23.6 &  9.7 & 13.9 &  0 & 16.5 &    8020 &  8.2 & \\
 WD\,J071839.44+520614.00    & 109.66381 &  52.10362 & cZZ    &  34.7 &  5.7 & 11.6 &  1 & 17.5 &   11330 &  8.0 & ZTF \\
 WD\,J073707.98+411227.88    & 114.28342 &  41.20747 & ZZ    &  25.1 & 10.1 & 15.0 &  0 & 15.8 &   11400 &  8.1 & \\
 WD\,J082309.66$-$015246.69  & 125.78997 &  -1.87977 & V     &  15.5 &  6.7 &  8.8 &  0 & 16.9 &   10790 &  7.7 & \\
 WD\,J082924.77$-$163337.25  & 127.35325 & -16.56033 & V     &  19.1 & 10.9 &  8.2 &  0 & 17.6 &   11000 &  8.0 & \\
 WD\,J084007.71+401503.73    & 130.03140 &  40.25101 & ZZ    &  13.1 &  6.7 &  6.4 &  0 & 15.7 &   11480 &  8.0 & \\
 WD\,J084652.90+442638.59    & 131.72060 &  44.44405 & ZZ    &  13.8 &  4.1 &  9.7 &  0 & 18.2 &   11250 &  8.0 & \\
 WD\,J085507.30+063541.14    & 133.78021 &   6.59446 & ZZ    &  15.2 &  6.5 &  8.7 &  0 & 17.3 &   10430 &  7.9 & \\
 WD\,J085722.50$-$224532.06  & 134.34310 & -22.75867 & V     &  16.3 & 12.6 &  3.8 &  0 & 15.2 &   11020 &  8.0 & \\
 WD\,J091635.08+385546.31    & 139.14593 &  38.92932 & ZZ    &  24.6 &  4.5 &  7.8 &  1 & 16.6 &   11700 &  8.1 & \\
 WD\,J091921.84$-$181719.09  & 139.84113 & -18.28863 & V     &  14.1 &  5.0 &  9.1 &  0 & 17.6 &   11620 &  8.0 & \\
 WD\,J092311.41+423634.16    & 140.79749 &  42.60934 & cTR   & 191.4 & 16.8 &  2.3 &  9 & 17.4 &   13110 &  8.0 & McD \\
 WD\,J092351.42+732624.15    & 140.96430 &  73.43996 & V     & 1615.5 & 55.0 & 45.9 & 15 & 18.4 &   15230 &  8.5 & \\
 WD\,J093250.57+554315.39    & 143.21045 &  55.72102 & V     &  31.0 &  2.3 & 13.1 &  1 & 17.7 &   11190 &  7.9 & \\
 WD\,J104233.54+405715.17    & 160.64015 &  40.95423 & ZZ    &  20.4 &  9.6 & 10.8 &  0 & 16.2 &   11160 &  8.0 & \\
 WD\,J104406.25+360046.79    & 161.02569 &  36.01279 & cZZ   &  24.1 &  6.9 &  5.2 &  1 & 17.6 &   11780 &  8.2 & McD                \\
 WD\,J105010.80$-$140436.76  & 162.54415 & -14.07693 & CV    &  73.1 & 14.5 & 22.0 &  1 & 17.2 &   10260 &  7.9 & \\
 WD\,J105256.27+130349.55    & 163.23401 &  13.06384 & V     &  17.2 &  7.3 &  9.9 &  0 & 17.4 &   10380 &  7.9 & \\
 WD\,J110121.95+401546.36    & 165.34128 &  40.26255 & V     &  73.7 &  9.5 &  8.9 &  3 & 17.5 &   10750 &  8.0 & \\
 WD\,J110604.54+180233.72    & 166.51879 &  18.04186 & cZZ   &  35.7 & 10.2 &  7.7 &  1 & 17.5 &   12610 &  8.8 & McD                \\
 WD\,J114833.63+012859.42    & 177.13994 &   1.48316 & TR    &  68.3 & 14.4 &  2.7 &  3 & 17.2 &   15310 &  8.1 & \\
 WD\,J115139.00+265413.46    & 177.91231 &  26.90366 & V     &  14.7 &  3.3 & 11.4 &  0 & 18.1 &   10900 &  7.9 & \\
 WD\,J120401.91$-$065328.98  & 181.00771 &  -6.89152 & V     &  22.5 &  5.1 &  6.2 &  1 & 17.3 &   11430 &  8.2 & \\
 WD\,J121924.49$-$083752.07  & 184.85179 &  -8.63114 & V     &  14.3 &  5.0 &  9.3 &  0 & 17.2 &   11080 &  7.8 & \\
 WD\,J121929.50+471522.94    & 184.87218 &  47.25636 & EB    &  17.9 &  1.7 &  7.2 &  1 & 17.6 &    7410 &  8.1 & \\
 WD\,J123446.89+164723.47    & 188.69500 &  16.78979 & V     &  22.8 &  4.4 &  7.0 &  1 & 18.2 &   11170 &  8.1 & \\
 WD\,J124804.03+282104.11    & 192.01683 &  28.35096 & V     &  15.1 &  4.7 & 10.4 &  0 & 18.0 &   11820 &  8.0 & \\
 WD\,J125009.02+771319.97    & 192.53681 &  77.22225 & V     &  15.7 &  6.9 &  1.0 &  1 & 16.3 &   15690 &  8.1 & \\
 WD\,J130110.52+010739.93    & 195.29393 &   1.12778 & ZZ    &  55.8 &  7.4 & 11.2 &  2 & 16.4 &   11090 &  7.9 & \\
 WD\,J130957.68+350947.19    & 197.48913 &  35.16320 & ZZ    &  13.8 &  5.8 &  8.0 &  0 & 15.3 &   11170 &  8.0 & \\
 WD\,J131246.44$-$232132.60  & 198.19216 & -23.35931 & CV    &  18.5 &  8.4 & 10.1 &  0 & 17.7 &    8820 &  8.1 & \\
 WD\,J132735.73+190345.90    & 201.89912 &  19.06256 & V     &  14.0 &  5.8 &  8.2 &  0 & 18.2 &   11380 &  7.9 & \\
 WD\,J134934.36+280948.93    & 207.39315 &  28.16362 & V     &  15.2 &  9.1 &  6.1 &  0 & 18.5 &   11240 &  8.1 & \\
 WD\,J135247.03$-$044137.04  & 208.19564 &  -4.69368 & V     &  14.0 &  6.6 &  7.4 &  0 & 17.4 &   11210 &  7.9 & \\
 WD\,J141126.22+200911.04    & 212.85905 &  20.15314 & EB    &  18.2 &  1.4 &  1.2 &  6 & 17.9 &   11480 &  7.9 & \\
 WD\,J141531.30+392357.57    & 213.87962 &  39.39944 & cZZ   &  29.7 &  6.3 &  8.6 &  1 & 17.1 &   11190 &  7.9 & McD                \\
 WD\,J141708.81+005827.32    & 214.28655 &   0.97425 & ZZ    &  13.7 &  3.5 & 10.2 &  0 & 18.2 &   10910 &  7.9 & \\
 WD\,J143047.25+510730.35    & 217.69666 &  51.12530 & cZZ     &  17.7 &  2.6 &  6.2 &  1 & 17.4 &   11000 &  8.0 & ZTF \\
 WD\,J144823.77+572454.62    & 222.09870 &  57.41506 & V     &  18.0 & 11.3 &  6.7 &  0 & 17.5 &   10730 &  7.9 & \\
 WD\,J145323.52+595056.24    & 223.34756 &  59.84885 & ZZ    &  18.9 &  2.9 &  6.6 &  1 & 17.2 &   11570 &  8.0 & \\
 WD\,J145540.89+175351.74    & 223.92038 &  17.89776 & V     &  31.7 &  2.1 &  8.4 &  2 & 17.4 &   10810 &  8.0 & \\
 WD\,J150549.19+110506.15    & 226.45496 &  11.08485 & V     &  15.1 &  3.5 & 11.6 &  0 & 18.2 &   10940 &  8.0 & \\
 WD\,J152809.27+553914.49    & 232.03818 &  55.65448 & cZZ     &  17.9 &  8.0 &  9.9 &  0 & 17.1 &   10290 &  7.5 & ZTF \\
 WD\,J153507.90+252118.56    & 233.78281 &  25.35498 & V     &  43.9 &  7.8 &  6.8 &  2 & 17.6 &   10380 &  8.0 & \\
 WD\,J153615.98$-$083907.53  & 234.06674 &  -8.65239 & CV    &  22.5 &  6.2 & 16.3 &  0 & 18.9 &    9280 &  8.0 & \\
 WD\,J154144.89+645352.98    & 235.43717 &  64.89778 & ZZ    &  84.6 & 10.5 & 17.7 &  2 & 15.6 &   11280 &  8.0 & \\
 WD\,J155129.23$-$191418.63  & 237.87131 & -19.23839 & V     & 322.4 & 19.5 & 45.0 &  4 & 18.1 &    7240 &  7.1 & \\
 WD\,J161042.95+503509.05    & 242.67887 &  50.58582 & V     &  26.1 &  2.9 & 10.1 &  1 & 18.1 &   10550 &  7.7 & \\
 WD\,J161903.72+091308.96    & 244.76549 &   9.21910 & V     &  12.9 &  9.7 &  3.2 &  0 & 18.2 &   11210 &  8.0 & \\
 WD\,J162436.81+321252.81    & 246.15377 &  32.21430 & cZZ   &  39.7 &  8.8 & 11.1 &  1 & 16.7 &   11200 &  7.9 & McD                \\
 WD\,J162813.34+122452.71    & 247.05510 &  12.41426 & ZZ    &  34.5 & 16.2 & 18.3 &  0 & 16.3 &   11410 &  8.0 & \\
 WD\,J163358.75+591206.59    & 248.49491 &  59.20201 & V     &  14.1 &  5.8 &  8.4 &  0 & 17.1 &   11240 &  8.0 & \\
 WD\,J163914.29+474835.84    & 249.80938 &  47.81017 & V     &  19.6 &  4.7 & 14.9 &  0 & 17.8 &   12090 &  8.1 & \\
 WD\,J170055.38+354951.09    & 255.23032 &  35.83057 & ZZ    &  66.8 &  6.3 & 15.9 &  2 & 17.4 &   11050 &  7.9 & \\
 WD\,J171251.78$-$191550.28  & 258.21559 & -19.26396 & cZZ   &  27.9 & 10.4 & 17.5 &  0 & 16.3 &   11910 &  8.1 & McD, CTIO               \\
 WD\,J171840.63+252431.90    & 259.66923 &  25.40876 & ZZ    & 218.8 &  7.1 & 14.8 &  9 & 16.1 &   11600 &  8.1 & McD                \\
 WD\,J174356.73+443852.18    & 265.98640 &  44.64807 & V     &  12.9 &  2.6 & 10.3 &  0 & 17.3 &   10620 &  7.9 & \\
 WD\,J175555.45+395844.68    & 268.98082 &  39.97867 & cZZ   & 114.6 &  8.5 & 10.6 &  5 & 17.0 &   11300 &  7.9 & McD                \\
 WD\,J181222.75+432107.27    & 273.09480 &  43.35229 & ZZ    &  14.8 &  6.3 &  8.5 &  0 & 16.3 &   11980 &  8.5 & McD                \\
 WD\,J181728.88+282326.60    & 274.37059 &  28.39079 & cZZ   &  72.9 &  3.2 & 11.4 &  4 & 18.0 &   10850 &  8.0 & McD                \\
 WD\,J182454.39+344331.58    & 276.22657 &  34.72561 & cZZ    &  15.1 &  1.5 &  6.0 &  1 & 18.3 &   11890 &  8.1 & ZTF \\
 WD\,J190024.55+191742.13    & 285.10238 &  19.29535 & cZZ   &  86.6 &  8.6 &  8.7 &  4 & 16.7 &   11560 &  8.0 & McD                \\
 WD\,J191852.42+583914.09    & 289.71819 &  58.65354 & cZZ   & 138.6 & 13.6 &  0.3 &  9 & 17.4 &   10750 &  8.0 & McD                \\
 WD\,J200736.50+174214.73    & 301.90242 &  17.70399 & CV    & 103.5 & 24.8 & 26.9 &  1 & 15.2 &    9760 &  7.7 & \\
 WD\,J202837.91$-$060842.77  & 307.15889 &  -6.14503 & cZZ   &  38.7 &  7.8 & 11.6 &  1 & 15.2 &   11820 &  8.5 & McD                \\
 WD\,J203724.74+570846.01    & 309.35311 &  57.14635 & V     &  27.6 &  6.0 &  7.8 &  1 & 17.2 &   12660 &  8.0 & \\
 WD\,J203737.79+323833.34    & 309.40762 &  32.64266 & cZZ   &  43.1 &  8.8 &  5.6 &  2 & 17.5 &   11480 &  7.5 & \\
 WD\,J204307.43+121926.72    & 310.78110 &  12.32407 & V     &  19.3 &  3.5 &  6.1 &  1 & 17.9 &   11800 &  8.1 & \\
 WD\,J205232.45$-$011244.62  & 313.13526 &  -1.21224 & cZZ   &  37.6 &  5.2 & 13.6 &  1 & 18.2 &   11270 &  7.9 & McD                \\
 WD\,J210452.73+233321.42    & 316.21954 &  23.55537 & ZZ    &  18.2 &  7.3 & 10.9 &  0 & 16.0 &   11250 &  8.0 & \\
 WD\,J215309.86+461902.53    & 328.29148 &  46.31748 & cZZ   &  21.2 &  2.4 &  8.3 &  1 & 17.8 &   11620 &  8.1 & McD                \\
 WD\,J220420.60+282321.36    & 331.08609 &  28.38937 & cZZ   &  18.1 &  1.9 &  7.2 &  1 & 17.8 &    9590 &  7.4 & McD                \\
 WD\,J221945.16$-$111142.54  & 334.93837 & -11.19500 & V     &  13.9 &  6.7 &  7.1 &  0 & 18.0 &   10850 &  8.0 & \\
 WD\,J225216.48$-$145907.61  & 343.06869 & -14.98539 & V     &  31.5 &  4.0 & 11.7 &  1 & 17.5 &   11310 &  8.0 & \\
 WD\,J230617.69+243207.68    & 346.57403 &  24.53522 & ZZ    &  24.8 & 10.8 & 14.0 &  0 & 15.4 &   11530 &  8.0 & \\
 WD\,J233401.45+392140.87    & 353.50572 &  39.36075 & CV    & 206.5 & 18.0 & 33.6 &  3 & 16.1 &    7700 &  7.0 & \\
 WD\,J233921.05+512410.79    & 354.83814 &  51.40253 & cZZ    &  18.9 &  8.3 & 10.5 &  0 & 16.8 &   11070 &  7.9 & ZTF \\
 WD\,J234830.32+415424.54    & 357.12632 &  41.90679 & V     &  21.3 &  3.8 &  6.8 &  1 & 18.1 &   11130 &  7.9 & \\
 WD\,J234949.49+225014.76    & 357.45643 &  22.83699 & cZZ   &  43.1 &  8.5 & 13.0 &  1 & 17.6 &   12130 &  8.1 & McD                \\
 WD\,J234953.46+130612.92    & 357.47404 &  13.10378 & cZZ   &  29.3 &  4.6 &  5.1 &  2 & 16.1 &   10660 &  7.9 & McD                \\
 WD\,J235010.39+201914.37    & 357.54317 &  20.32053 & cZZ    &  15.2 &  2.3 & 12.9 &  0 & 17.4 &   11870 &  8.0 & ZTF \\
 WD\,J235915.98+653225.97    & 359.81627 &  65.54060 & V     &  18.8 &  4.3 &  5.1 &  1 & 18.0 &   12330 &  8.1 & \\
\enddata
\end{deluxetable}
\onecolumngrid

%%%%%%%%%%%%%%%%%%%%%%%%%%%%%%%%%%%%%%%%%%%%%%%%%%%%%%%%%%%%%%%%%%%%%%%%%%%%%%%

\section{Table of Pulsation Spectra for Newly Confirmed ZZ Cetis with Follow-up Photometry}\label{sec:pulsation_spectra}

Motivated by \citealt{2006ApJ...640..956M}, we catalog here in Table~\ref{tab:pulsations} the linearly independent peaks greater than the $4\langle A \rangle$ significance threshold from our follow-up high-speed photometric observations using the 2.1\,m Otto Struve telescope at McDonald Observatory in the {\em BG40} bandpass filter (unless specified otherwise) and the 0.9\,m SMARTS telescope at Cerro Telolo Inter-American Observatory in the $V$-band. Linearly independent peaks were determined by using the prewhitening tools provided by the \textsc{pyriod} Python package. We present the periodicities in descending order by amplitude. The weighted mean period (WMP) of these pulsations for each object is also presented.

We also report all the linearly independent periods in the ZTF time-series with amplitudes greater than the 0.1\% false alarm probability significance thresholds we estimated from our bootstrapping exercise for the objects in the top 1\%. We distinguish these periodicities by the ZTF light curves: $g$, $r$, and the combined $g$+$r$ light curve. While most of these periods are characteristic of ZZ Ceti pulsations, we denote objects that are known non-pulsators or whose periodicities are not indicative of pulsations with a $\ddagger$ in the WMP column to avoid confusion.

\startlongtable
\begin{deluxetable}{ccccl}
\tablecolumns{5}
\tablenum{6}
\tablewidth{0pt}
\tabletypesize{\small}
\tablecaption{Pulsation Spectra for newly confirmed ZZ Cetis.}
\label{tab:pulsations}
\tablehead{
\colhead{WD} & \colhead{$T_{\mathrm{eff}}$} & \colhead{$\log(g)$} & \colhead{WMP} & \colhead{Linearly Independent Modes: Period/Amplitude}\\
\colhead{} & \colhead{(K)} & \colhead{(cgs)} & \colhead{(sec)} & \colhead{(sec) / (\%) }
}
\startdata
\cutinheadnew{McDonald 2.1\,m Pulsators}
 WD\,J001038.25+173907.24 &   11220 &   7.9 &  855.9 & 978.2/3.4, 702.0/2.7 \\
 WD\,J023610.51+223431.86 &    9660 &   7.7 &  812.4 & 831.7/4.3, 898.5/3.2, 618.6/1.5, 579.2/1.1, 990.9/1.0 \\
 WD\,J024029.66+663637.06 &   11870 &   8.0 &  411.2 & 864.5/2.0, 982.9/1.0, 1058.2/0.9, 623.9/0.4 \\
 WD\,J032438.66+602055.88 &   11250 &   8.0 &  797.5 & 818.3/2.0, 894.1/0.8, 691.8/0.7, 999.5/0.7, 506.7/0.6 \\
 WD\,J043350.99+485039.18 &   11370 &   8.0 &  549.5 & 510.6/3.1, 583.7/2.8, 464.2/1.6, 667.2/1.4 \\
 WD\,J044832.11$-$105349.85 &   12250 &   8.5 &  508.0 & 536.4/5.3, 316.8/0.8 \\
 WD\,J045132.19$-$033308.43 &   11290 &   8.0 &  839.5 & 831.3/3.0, 908.8/2.3, 751.7/1.5 \\
 WD\,J060102.01+541757.82 &   11110 &   8.1 &  861.5 & 861.5/0.3 \\
 WD\,J062154.74$-$021605.36$^{\dagger}$ &    9490 &   7.6 &  767.8 & 788.5/2.0, 874.9/1.8, 645.5/1.6, 510.5/1.5, 1095.5/1.0 \\
 WD\,J104406.25+360046.79 &   11780 &   8.2 &  628.6 & 661.6/1.6, 361.8/1.5, 897.5/1.3 \\
 WD\,J110604.54+180233.72 &   12610 &   8.8 &  417.4 & 369.7/10.2, 1180.2/0.6, 537.2/0.5 \\
 WD\,J141531.30+392357.57 &   11190 &   7.9 &  786.5 & 953.6/4.3, 735.3/2.9, 710.8/2.6, 399.6/1.0 \\
 WD\,J162436.81+321252.81 &   11200 &   7.9 &  845.6 & 930.2/6.0, 648.1/2.6 \\
 WD\,J171251.78$-$191550.28 &   11910 &   8.1 &  574.2 & 569.9/5.4, 488.7/3.1, 762.7/2.3, 455.0/1.2 \\
 WD\,J171840.63+252431.90 &   11600 &   8.1 &  481.1 & 494.0/3.6, 397.3/1.2, 531.9/1.0 \\
 WD\,J175555.45+395844.68 &   11300 &   7.9 &  780.7 & 920.8/2.5, 530.5/2.2, 656.5/1.7, 1286.8/0.8 \\
 WD\,J181222.75+432107.27 &   11980 &   8.5 &  556.8 & 407.2/3.7, 788.7/1.7, 741.4/1.4, 476.6/1.2 \\
 WD\,J181728.88+282326.60 &   10850 &   8.0 &  614.2 & 581.4/4.0, 464.9/1.1, 747.2/1.0, 811.4/0.8 \\
 WD\,J184424.33+504727.95 & $*$ & $*$ & 3700.6 & 4188.2/6.4, 2922.9/1.7, 1863.2/1.0 \\
 WD\,J190024.55+191742.13 &   11560 &   8.0 &  483.8 & 507.2/4.1, 546.3/1.7, 285.6/0.8, 401.3/0.6 \\
 WD\,J191852.42+583914.09 &   10750 &   8.0 & 1192.8 & 1099.7/0.9, 997.3/0.8, 1497.3/0.4, 861.3/0.4, 1952.8/0.3 \\
 WD\,J202837.91$-$060842.77 &   11820 &   8.5 &  696.2 & 526.5/1.9, 412.6/1.7, 703.9/1.3, 1987.2/0.7, 587.8/0.4 \\
 WD\,J203737.79+323833.34 &   11480 &   7.5 & 1326.4 & 1296.2/4.3, 1016.3/3.2, 1661.5/2.7, 1769.8/2.5, 1126.5/1.7, 922.4/1.4 \\
 WD\,J205232.45$-$011244.62 &   11270 &   7.9 &  784.4 & 882.4/7.0, 626.5/1.9, 518.1/1.4 \\
 WD\,J215309.86+461902.53 &   11620 &   8.1 &  719.3 & 646.9/2.5, 907.3/2.5, 830.1/1.7, 502.9/1.4, 586.1/1.3 \\
 WD\,J220420.60+282321.36 &    9590 &   7.4 &  735.5 & 532.1/2.8, 932.3/2.2, 566.7/1.6, 476.1/0.8, 1277.5/0.7, 1172.0/0.5 \\
 WD\,J234953.46+130612.92 &   10660 &   7.9 & 1049.7 & 1078.1/1.9, 1166.5/1.4, 990.4/0.9, 775.9/0.9, 1533.6/0.6, 834.2/0.4, 739.5/0.4 \\
 WD\,J234949.49+225014.76 &   12130 &   8.1 &  867.4 & 872.7/9.4, 532.8/2.0, 1265.4/1.5 \\
\hline
\cutinheadnew{CTIO 0.9\,m Pulsators}
 WD\,J100223.96$-$575507.91 &   12110 &   8.0 &  747.0 & 708.5/2.5, 805.4/1.6 \\
 WD\,J111221.43$-$513003.90 &   11190 &   8.0 &  788.7 & 764.9/3.9, 828.0/2.3 \\
 WD\,J115414.55$-$592658.81 &   11170 &   8.0 &  890.5 & 890.5/3.5 \\
 WD\,J171251.78$-$191550.28 &   11910 &   8.1 &  522.8 & 570.4/4.3, 490.5/3.3, 455.0/1.4 \\
 WD\,J174349.28$-$390825.95 &   11560 &   8.1 &  763.0 & 785.3/3.0, 605.7/2.5, 871.7/2.2, 732.1/2.0, 488.9/1.4, 680.9/1.4, 1492.1/0.9 \\
 WD\,J204349.21$-$390318.02 &   11270 &   8.0 &  773.4 & 857.5/1.9, 940.3/0.9, 694.2/0.5, 486.4/0.4, 366.4/0.4 \\
 WD\,J211815.52$-$531322.72 &   11120 &   7.9 &  943.4 & 943.4/2.4 \\
 WD\,J224840.07$-$064244.65 &   11640 &   8.1 &  529.8 & 558.3/3.7, 481.7/2.2 \\
 \hline
\cutinheadnew{ZTF Significant Periods}
 WD\,J004711.37+305609.18 &   10450 &   7.5 & 813.1 & $r$: 940.9/2.1; $g$+$r$: 940.9/2.2, 618.7/1.7 (4.66$\langle A \rangle$) \\
 WD\,J010528.63+020501.63 &   11180 &   7.9 & 453.6 & $g$+$r$: 453.6/2.0 (4.96$\langle A \rangle$) \\
 WD\,J014635.73+491443.10 &    8550 &   8.3 & $\ddagger$ & $r$: 7407.7/8.5; $g$+$r$: 7407.7/4.9 (5.07$\langle A \rangle$) \\
 WD\,J024029.66+663637.06 &   11870 &   8.0 & 542.3 & $g$+$r$: 542.3/1.6 (4.96$\langle A \rangle$) \\
 WD\,J041856.64+271748.31 &   12380 &   7.8 & 516.9 & $g$+$r$: 537.7/2.8, 494.2/2.1 (4.63$\langle A \rangle$) \\
 WD\,J053432.93+770757.40 &   10410 &   8.3 & $\ddagger$ & $g$: 2604.5/7.6, 1302.2/1.8; $r$: 2604.5/3.6, 1302.2/1.6; \\
   &  &  &  & $g$+$r$: 2604.5/5.2, 1302.2/1.6 (5.01$\langle A \rangle$) \\
 WD\,J062516.34+145558.50 &   11280 &   8.1 & 831.5 & $r$: 836.6/2.4, 610.6/1.9; $g$+$r$: 832.4/1.8, 582.0/1.5, 979.1/1.4 (4.34$\langle A \rangle$) \\
 WD\,J062555.04-141442.31 &    8020 &   8.2 & $\ddagger$ & $g$: 13542.1/5.5; $r$: 13541.9/4.6; $g$+$r$: 13542.0/5.0 (4.86$\langle A \rangle$) \\
 WD\,J071839.44+520614.00 &   11330 &   8.0 & 627.4 & $g$+$r$: 627.4/2.2 (4.95$\langle A \rangle$) \\
 WD\,J091635.08+385546.31 &   11700 &   8.1 & 476.1 & $g$+$r$: 498.1/2.2, 451.9/1.7 (5.00$\langle A \rangle$) \\
 WD\,J121929.50+471522.94 &    7410 &   8.1 & $\ddagger$ & $g$: 54949.0/3.0; $r$: 54959.6/2.5; $g$+$r$: 54953.6/2.6 (5.11$\langle A \rangle$) \\
 WD\,J141531.30+392357.57 &   11190 &   7.9 & 1002.9 & $g$+$r$: 1002.9/1.2 (4.59$\langle A \rangle$) \\
 WD\,J145323.52+595056.24 &   11570 &   8.0 & 458.1 & $g$: 458.1/3.0; $r$: 458.1/2.5; $g$+$r$: 458.1/2.8 (5.02$\langle A \rangle$) \\
 WD\,J152809.27+553914.49 &   10290 &   7.5 & 769.1 & $g$: 769.1/2.7; $r$: 769.1/1.9; $g$+$r$: 769.1/2.1 (5.09$\langle A \rangle$) \\
 WD\,J170055.38+354951.09 &   11050 &   7.9 & 696.5 & $g$: 508.0/2.1, 558.8/1.8; $r$: 555.2/1.5, 646.3/1.2, 514.1/1.1; \\
   &  &  &  & $g$+$r$: 555.2/1.6, 508.0/1.6, 891.8/1.0 (4.33$\langle A \rangle$) \\
 WD\,J171251.78-191550.28 &   11910 &   8.1 & 528.3 & $r$: 573.8/3.9, 488.6/2.6, 261.5/1.4, 287.9/1.3, 746.1/1.1, 447.2/1.1; \\
   &  &  &  & $g$+$r$: 573.8/3.8, 485.8/2.5, 259.1/1.5, 449.5/1.2, 752.7/1.2, 290.7/1.1 (4.07$\langle A \rangle$) \\
 WD\,J171840.63+252431.90 &   11600 &   8.1 & 481.7 & $g$: 495.2/2.3, 397.5/1.4, 532.1/1.1; $r$: 495.2/1.8, 401.2/0.9; \\
   &  &  &  & $g$+$r$: 495.2/2.0, 397.5/1.2, 532.1/0.9 (4.59$\langle A \rangle$) \\
 WD\,J175555.45+395844.68 &   11300 &   7.9 & 701.2 & $g$+$r$: 701.2/1.4 (5.06$\langle A \rangle$) \\
 WD\,J181222.75+432107.27 &   11980 &   8.5 & 353.4 & $g$+$r$: 353.4/1.2 (4.96$\langle A \rangle$) \\
 WD\,J181728.88+282326.60 &   10850 &   8.0 & 608.9 & $g$+$r$: 585.0/1.8, 631.0/1.7 (4.89$\langle A \rangle$) \\
 WD\,J190024.55+191742.13 &   11560 &   8.0 & 510.4 & $g$: 510.4/3.1; $g$+$r$: 510.4/2.3 (4.90$\langle A \rangle$) \\
 WD\,J200736.50+174214.73 &    9760 &   7.7 & $\ddagger$ & $g$: 2520.5/6.3; $r$: 2448.9/6.0, 64954.5/3.4, 1569.9/2.5; \\
   &  &  &  & $g$+$r$: 2448.9/6.1, 263876.6/3.0, 1540.5/2.2 (5.20$\langle A \rangle$) \\
 WD\,J215309.86+461902.53 &   11620 &   8.1 & 942.9 & $r$: 921.1/1.7; $g$+$r$: 942.9/1.4 (5.02$\langle A \rangle$) \\
 WD\,J233401.45+392140.87 &    7700 &   7.0 & $\ddagger$ & $g$: 2432.6/4.0; $r$: 4823.2/3.7, 2432.6/3.3; $g$+$r$: 2432.6/3.7, 4823.2/2.8 (4.92$\langle A \rangle$) \\
 WD\,J233921.05+512410.79 &   11070 &   7.9 & 1050.6 & $r$: 1037.4/1.6; $g$+$r$: 1050.6/1.3 (4.74$\langle A \rangle$) \\
 WD\,J235010.39+201914.37 &   11870 &   8.0 & 364.2 & $g$: 365.7/2.4; $r$: 364.2/1.5; $g$+$r$: 364.2/1.8 (4.84$\langle A \rangle$) \\
\enddata
\tablenotetext{*}{$\,\,-$ The lowest-mass ELM white dwarfs do not have $T_\mathrm{eff}$ and $\log(g)$ values in the \citet{GentileFusillo2019} {\em Gaia} DR2 catalog.}
\tablenotetext{\dagger}{$\,\,-$ Observed in SDSS $g$-band}
\tablenotetext{\ddagger}{$\,\,-$ The weighted mean period of pulsations is only relevant to pulsating white dwarfs.}
\end{deluxetable}
\onecolumngrid

\end{document}